\documentclass[journal]{IEEEtran}

\usepackage{cite}
\usepackage{amsmath,amsfonts}
\usepackage{algorithmic}
\usepackage{array}
\usepackage[caption=false,font=normalsize,labelfont=sf,textfont=sf]{subfig}
\usepackage{textcomp}
\usepackage{stfloats}
\usepackage{url}
\usepackage{verbatim}
\usepackage{graphicx}
\def\BibTeX{{\rm B\kern-.05em{\sc i\kern-.025em b}\kern-.08em
    T\kern-.1667em\lower.7ex\hbox{E}\kern-.125emX}}
\usepackage{balance}

\usepackage{multicol}
\usepackage{xcolor}
\usepackage{hyperref}

\def\BibTeX{{\rm B\kern-.05em{\sc i\kern-.025em b}\kern-.08em
    T\kern-.1667em\lower.7ex\hbox{E}\kern-.125emX}}
\begin{document}

\title{Visual Analytics of Multivariate Networks with \\Representation Learning and\\Composite Variable Construction
}

\author{Hsiao-Ying Lu, Takanori Fujiwara, Ming-Yi Chang, Yang-chih Fu, Anders Ynnerman, and Kwan-Liu Ma
\thanks{Hsiao-Ying Lu and Kwan-Liu Ma are with University of California, Davis. E-mail: \{hyllu, klma\}@ucdavis.edu.}
\thanks{Takanori Fujiwara and Anders Ynnerman are with Link\"oping University. E-mail: \{takanori.fujiwara, anders.ynnerman\}@liu.se.}
\thanks{Ming-Yi Chang is with Fu Jen Catholic University. E-mail: mychang@mail.fju.edu.tw.}
\thanks{Yang-chih Fu is with Academia Sinica. E-mail:fuyc@gate.sinica.edu.tw.}
}

\markboth{IEEE Transactions on Visualization and Computer Graphics, VOL. X, NO. X, MONTH 2024}
{H.-Y. Lu,
\MakeLowercase{\textit{(et al.)}:
Visual Analytics of Multivariate Networks with Representation Learning and Composite Variable Construction}}

\maketitle

\begin{abstract}
Multivariate networks are commonly found in real-world data-driven applications. Uncovering and understanding the relations of interest in multivariate networks is not a trivial task. This paper presents a visual analytics workflow for studying multivariate networks to extract associations between different structural and semantic characteristics of the networks (e.g., what are the combinations of attributes largely relating to the density of a social network?). The workflow consists of a neural-network-based learning phase to classify the data based on the chosen input and output attributes, a dimensionality reduction and optimization phase to produce a simplified set of results for examination, and finally an interpreting phase conducted by the user through an interactive visualization interface. A key part of our design is a composite variable construction step that remodels nonlinear features obtained by neural networks into linear features that are intuitive to interpret. We demonstrate the capabilities of this workflow with multiple case studies on networks derived from social media usage and also evaluate the workflow with qualitative feedback from experts.
\end{abstract}


\newcommand{\ProbDensFunc}{f}
\newcommand{\ProbDensFuncAll}{\ProbDensFunc_\mathrm{all}}
\newcommand{\ProbDensFuncClassZero}{\ProbDensFunc_\mathrm{0}}
\newcommand{\ProbDensFuncClassOne}{\ProbDensFunc_\mathrm{1}}

\newcommand{\nInsts}{n}
\newcommand{\nInstsClassZero}{\nInsts_\mathrm{0}}
\newcommand{\nInstsClassOne}{\nInsts_\mathrm{1}}

\newcommand{\DensRatioFunc}{g}
\newcommand{\DensRatioFuncClassZero}{\DensRatioFunc_\mathrm{0}}
\newcommand{\DensRatioFuncClassOne}{\DensRatioFunc_\mathrm{1}}

\newcommand{\Point}{\mathbf{p}}

\newcommand{\TransRate}{r}
\newcommand{\CaptionWorkflow}{The visual analytics workflow for investigating associations in multivariate networks, where Steps 1--4 are executed with a script for machine learning and Steps 5--6 are conducted interactively with our UI.}

\newcommand{\CaptionUI}{The visual interface for facilitating interactive analysis using the workflow in \autoref{fig:workflow}. 
(a) Relating to Step 4, this view visualizes input attributes' contributions to the 1D network representation. 
(b) Composite variables generated in Step 5 are shown as a list of scatterplots, where the 1D network representation and composite variable correspond to $x$- and $y$-axes, respectively (see b2). 
As shown in b1, when a composite variable is not generated yet, a swarm-plot-like visualization presents the 1D network representation generated through Step 3 to help assess its quality.
For a and b, we employ our two-class density scatterplots.
(c) A node-link diagram and (d) a set of histograms inform the network structure and attribute distributions.
(e) Other auxiliary information is displayed, including the prediction accuracy of NNs trained for Step 2.}

\newcommand{\CaptionSHAP}{Comparison of SHAP value visualizations: (a) the default plot in the SHAP package~\cite{shap_library}; (b) the plot colored by class; and (c) our design.}

\newcommand{\CaptionScatterplots}{The comparison of scatterplot designs: (a) scatterplot with colored classes, (b) density scatterplot, (c,d) scatterplots encoding the total density and the ratio of each class's density with two different bivariate colormaps. (d) is our final design for a two-class density scatterplot.}

\newcommand{\CaptionCaseOne}{Study 1: Networks after selecting students with (a1) low (--\,0.9\,\texttt{open} +\,0.4\,\texttt{extra}) and (a2) high \texttt{extra}; distributions of \texttt{neuro} (b1, b2) corresponding to the selections of (a1, a2).}

\newcommand{\CaptionCaseTwo}{Study 2-1: The comparison of \texttt{business} and \texttt{engineering} students: (a) the SHAP values and (b-e) composite variables constructed from network centralities. Combining \texttt{eigenvector} and \texttt{total\_degree} significantly increases the correlation.}

\newcommand{\CaptionCaseTwoCont}{Study 2-2: The comparison of the high and low \texttt{eigenvector} groups of \texttt{engineering} students: the SHAP value visualizations (a) before and (b) after improving the feature extraction setting.}

\newcommand{\CaptionCaseThree}{Study 3: (a) The top contributing attributes and composite variables constructed with (b) the top 3 and (c) top 5 attributes.}
\section{Introduction}
\IEEEPARstart{M}{ultivariate} networks~\cite{kerren2014multivariate} consisting of both topological and semantic information can model complex relations between entities.
One common analysis task performed on multivariate networks is to understand associations among structural and semantic characteristics~\cite{kerren2014multivariate,atzmueller2021mining,chetty2022social,chen2023calliopenet}.
For example, from social media usage, analysts may want to see how likely each possible combination of individual characteristics (e.g., age, gender, and extraversion) and friendship structures is related to their addiction level to social media. 
Such analysis can be more complicated when the associations underlie intertwining factors.

To aid in analysis of multivariate networks, researchers have introduced visual analytic support, including network layouts considering semantic information, interactive simplification, and incorporation of coordinated views~\cite{kerren2014multivariate,nobre2019state}. 
Among others, utilizing network representation learning (NRL)~\cite{zhang2018network,huang2023va} is one promising approach as it can capture latent features of networks and highlight essential aspects that should be examined with visualizations. 
Existing visual analytics methods~\cite{fujiwara2022network,fujiwara2020visual,martins2012multidimensional,martins2017mvn,vandenelzen2016reducing,song2022interactive} utilize NRL methods to learn networks' \textit{general} representations---overall summaries of networks (e.g., variance of node degrees).
However, to identify associations, a general-purpose NRL is not effective as it is not particularly designed to capture latent features important for associations of interest.

On the other hand, as seen in the machine learning (ML) field, NRL using neural networks (NNs) can learn representations specific to an analysis focus by using appropriate loss functions~\cite{hamilton2017inductive}.
While these representations are suitable for fully automated analyses, such as network classification, it is often difficult for analysts to interpret the extracted features.
This is not preferable when analysts want to be involved in the analysis process and derive insights into associations.

To address the aforementioned problems, we introduce a visual analytics workflow that provides (1) network representations specific to the structural and semantic associations in multivariate networks as well as (2) interpretation supports for the analysis results.

To learn such representations, we first extract essential structural features by using an NRL method and then train an NN model that is designed to classify values of a user-selected attribute (e.g., high and low addiction levels to social media).
After training, the model generates latent features that are highly related to the selected attribute. 

For the interpretation of the obtained network representations, we employ dimensionality reduction (DR), interactive visualization, and composite variables~\cite{song2013composite}.
To help analysts assess the quality of the representations, we simplify the latent features with a linear DR method (specifically, linear discriminant analysis or LDA).
We then visualize the information of the simplified features in a 2D plot, where the distribution of networks or their elements (i.e., nodes and links) is shown along a latent direction that is related to the classification.
From this plot, analysts can judge which part of networks or elements (e.g., subjects with age 20--30) likely holds clearer associations with the selected attribute.
This analysis is almost infeasible if we rely only on the classification quality measures. 
In addition, we introduce a mechanism of composite variable construction to explain the meanings of the network representations.
The mechanism first suggests network structures and attributes highly related to the representations by utilizing a model-agnostic interpretation method, the SHAP~\cite{lundberg2017shap}.
Then, after analysts interactively select structures and attributes from the suggestion, an optimization method automatically generates a composite variable that resembles the network representation.
By examining this composite variable, analysts can understand the associations among the selected attributes and the other information.

To support effective analysis of the associations, we develop a new density scatterplot, named \textit{two-class density scatterplot}, that can depict a trend of data distribution as well as inform the mixture and separation of two classes even from numerous instances (e.g., 1000 instances).
Incorporating two-class density scatterplots, we develop an interactive interface that links visualizations designed to enhance the visual analytics workflow.
We further demonstrate the capabilities of the workflow and interface with three case studies using two real-world datasets on social media usage. 
Moreover, we validate the usability of the workflow through experts' feedback.

In summary, we consider our primary contributions to be:
\begin{itemize}
    \item a visual analytics workflow considering both generation and interpretation of network representations that are expressive for user-specified analysis targets;
    \item a workflow of composite variable construction that aids in attribute selection and attribute weight optimization for interpretations with an influence from multiple attributes;
    \item a two-class density scatterplot as a versatile visualization to review individual- and group-level data patterns with class information.
\end{itemize}
We provide source code related to the workflow and a demonstration video of interactive analysis~\cite{supp}.
\section{Related Work}
\label{sec:related_work}

\noindent
Our work is closely related to two topics in the visualization field: NRL and the interpretation of learned representations. 
For the broader discussion on visualizations for analyzing networks and interpreting ML results, refer to existing surveys~\cite{kerren2014multivariate,mcgee2019state,beck2017taxonomy,chatzimparmpas2020state}.

\subsection{Learning Network Representation}

\noindent
NRL aims to generate a set of low-dimensional vectors (also called representation) that captures certain important characteristics of networks, nodes, or links~\cite{zhang2018network}. 
The representation is usually learned for downstream tasks, such as node classification and link prediction. 
Various NRL methods are developed, including node2vec~\cite{grover2016node2vec}, graph convolutional networks~\cite{kipf2016semi}, graph neural networks (GNNs) with the self-attention~\cite{ying2021transformers}, to name but a few~\cite{zhang2018network}. 

The researchers have been utilizing NRL for visualization to interactively examine complex network datasets.
For example, Freire et al.~\cite{freire2010manynets} represented one network by a set of network statistics (e.g., degree distribution) to compare many networks in a tabular interface. 
Gove~\cite{gove2019gragnostics} suggested several network-level features (e.g., density) that are easier to interpret and faster to compute for interactive visualization.
Other researchers utilized the occurrences of graphlets~\cite{prvzulj2007biological} (small, connected, non-isomorphic subgraph patterns) to identify visually similar networks~\cite{von2009visual,harrigan2012egonav,kwon2017would}.
When node correspondence exists among networks, another common approach is directly applying DR methods to networks' adjacency matrices to capture the similarities of networks~\cite{bach2016time,fujiwara2017visual}.
Van den Elzen et al.~\cite{vandenelzen2016reducing} took a similar DR approach while further incorporating network statistics. 
Martins et al.~\cite{martins2012multidimensional,martins2017mvn} used DR to lay out nodes by their structural and semantic similarities.

Similar to ours, recently, a few works employed NN-based NRL.
Fujiwara et al.~\cite{fujiwara2022network} introduced contrastive NRL (cNRL) by integrating a variant of GNNs and contrastive learning~\cite{zou2013contrastive}.
cNRL extracts a representation of two networks to highlight salient characteristics in one network relative to another.
Utilizing linear DR, they further designed an interpretable cNRL method and enhanced it with interactive visualizations~\cite{fujiwara2020visual}. 
Song et al.~\cite{song2022interactive} used GNNs to support interactive subgraph pattern search, where GNNs are used to covert each network in a comparable, fixed-length latent vector. 

Unlike the above approaches, we use NN-based NRL to obtain representations that are specifically for uncovering the associations of interest in multivariate networks.
Also, we address the interpretation of network representations with composite variable construction, which is easier to understand when compared with the approaches referring to complex coefficients in linear DR results~\cite{fujiwara2020visual,fujiwara2022network}.

\subsection{Interpreting Representations}

\noindent
Although the interpretation support for NRL is still sparsely studied (e.g., \cite{fujiwara2020visual}), a variety of interpretation methods are developed to explain high-dimensional data representations that are extracted by DR methods or ML models~\cite{molnar2020interpretable, huang2023va, la2023state, choo2018visual, hohman2018visual}. 
Existing interpretation methods can be mainly categorized into two approaches: (1) identifying essential information to specific patterns found in representations (i.e., post-hoc explanation approach) and (2) constructing simple, interpretable representations during a learning phase (i.e., explainability-by-design approach~\cite{hamon2020robustness}).

\subsubsection{Post-Hoc Approach}

\noindent
Within visual analytics methods, researchers have identified influential attributes on the cluster formation in DR results from statistical charts (e.g., boxplots of attributes for each cluster)~\cite{kwon2018clustervision,neto2021multivariate,vanozenoodt2022outoftheplane}.
As univariate statistics are often insufficient to capture the cluster characteristics, researchers further considered influences from multiple attributes~\cite{fujiwara2020supporting,joia2015uncovering,turkay2012representative,zhou2016dimension,zang2022evnet}.
For example, Joia et al.~\cite{joia2015uncovering} applied PCA to each cluster to examine multivariate influences.

Another common visual analytics strategy is utilizing measures that inform how strongly the change of an input attribute value influences an ML inference result (e.g., how many more bikes are rented if the temperature increases one degree)~\cite{krause2016interacting, wexler2019if, apley2020visualizing, li2020multimodel, angelini2023visual}. 
Such measures include partial dependence (PD) and SHAP values~\cite{lundberg2017shap}.
For example, Krause et al.~\cite{krause2016interacting} developed a system to interactively investigate PD. 
Their system only supports analysis of PD that is obtained by changing each attribute's values individually---interpretation only from a single attribute level. 
Angelini et al.~\cite{angelini2023visual} introduced a visual analytics framework that can even deal with PD derived from a simultaneous change of multiple attributes.

Similar to Angelini et al.'s work~\cite{angelini2023visual}, composite variable construction in our workflow also allows a multi-attribute level interpretation.
However, PD or other perturbation-based measures (e.g., SHAP values) merely output a value for each perturbed input.
Consequently, it is not trivial to induce numerical relationships between input attributes and ML inference results. 
In contrast, our workflow's composite variable provides a linear function that summarizes the associations among multiple attributes.
This ability helps us perform the interpretation both efficiently and intuitively.

\subsubsection{Explainability-By-Design Approach} 
When compared with the post-hoc approach, the explainability-by-design approach is taken by a smaller number of visual analytics works.
Knittle et al.~\cite{knittel2020visual} used NNs consisting of one hidden layer with a small number of NN nodes to extract nonlinear representations that relate input attributes to a target output attribute.
This simple NN allows analysts to identify 
a small number of representations that show clear associations between the input and target attributes. 
Gleicher~\cite{gleicher2013explainers} produced simple composite variables that are to classify a user-selected attribute.
To craft such composite variables, Gleicher performed a support-vector machine-based exhaustive search for the selection of composing variables while considering a balance between their simplicity and expressiveness.

In terms of using NNs to extract the input-output relationships, the work by Knittle et al.~\cite{knittel2020visual} is closely related to ours. 
However, their interpretation of the obtained representations is based only on univariate value distributions, which is insufficient when NNs capture complex input-output relationships.
Similar to Gleicher's work~\cite{gleicher2013explainers}, our work crafts simple composite variables, but we do not involve the computationally expensive exhaustive search. 
Instead, we rank variables based on their contributions to the NNs' predictions and involve analysts' knowledge to select attributes of interest.

\section{Methodology}

\noindent
This section introduces our visual analytics workflow and two-class density scatterplot.
We first describe design considerations of the workflow, which we identify based on our literature survey in \autoref{sec:related_work} and targeted analysis on multivariate networks.
We then provide an overview of the workflow, followed by the details of each step.

\subsection{Design Considerations}

\noindent
The following design considerations (DCs) are identified to support the task of understanding associations of interest and fill the analytical gap that is not covered by existing methods.

\newcommand{\Flexibility}{DC1: Flexibility}
\newcommand{\Expressivity}{DC2: Expressivity}
\newcommand{\Interpretability}{DC3: Interpretability}
\newcommand{\Steerability}{DC4: Tunability}
\newcommand{\Extensibility}{DC5: Extensibility}

\textbf{\Flexibility.} Analysis of multivariate networks requires reviewing intertwined relationships in both structural and semantic information~\cite{kerren2014multivariate}. 
Also, based on analysis purposes and available datasets, analysts often want to investigate networks from different levels, such as node, link, and network levels (e.g., node: \cite{martins2017mvn,fujiwara2020visual}; 
link: \cite{crnovrsanin2014visualization};
and network levels: \cite{song2022interactive,freire2010manynets,kwon2017would,vandenelzen2016reducing}).
The workflow should provide flexibility to conduct analyses from multiple aspects at various levels.

\textbf{\Expressivity.} 
The main objective of our work is to support examining associations among target (e.g., the addiction level to social media) and other related information (e.g., gender, age, and friendships). 
The workflow should be able to extract features that are expressive for this analysis task (i.e., containing important information to reveal associations) rather than providing a general summary of networks. 

\textbf{\Interpretability.} Interpretation is essential to gain insights into analysis results~\cite{knittel2020visual,gleicher2013explainers}. 
Because expressive features such as those extracted by NNs are often difficult to interpret only from each individual attribute~\cite{fujiwara2020visual,cunningham2014dimensionality}, the workflow should provide more advanced interpretation support that considers the combined influence from multiple attributes.

\textbf{\Steerability.}
Conflict often exists between the expressiveness and interpretability of features (i.e., more expressive, more difficult to interpret)~\cite{gleicher2013explainers,zou2006sparse,malerba1996further}.
Also, the required preciseness of interpretation can be varied by analysis (i.e., complicated but precise interpretation vs. less precise but simpler interpretation). 
The workflow should allow control of the balance of expressiveness and interoperability.
    
\textbf{\Extensibility.} Although we provide a concretized method for each workflow step, a different set of methods may be more suitable according to the characteristics of future datasets and analysis goals~\cite{cunningham2015linear,wu2020comprehensive,zhang2018network,fujiwara2021interactive}.
The workflow should be extensible to incorporate new or different methods.

\begin{figure*}[t]
    \centering
    \includegraphics[width=0.9\linewidth]{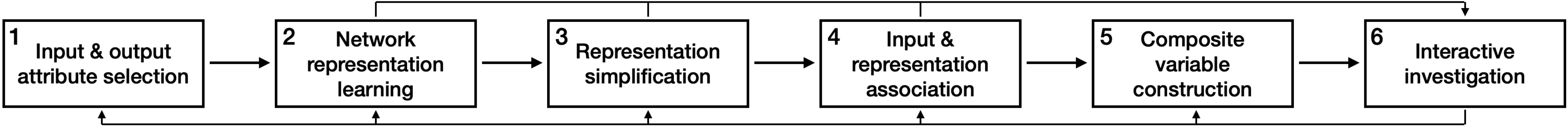}
    \caption{\CaptionWorkflow{}}
    \label{fig:workflow}
\end{figure*}
\begin{figure*}[t]
    \centering
    \includegraphics[width=\linewidth]{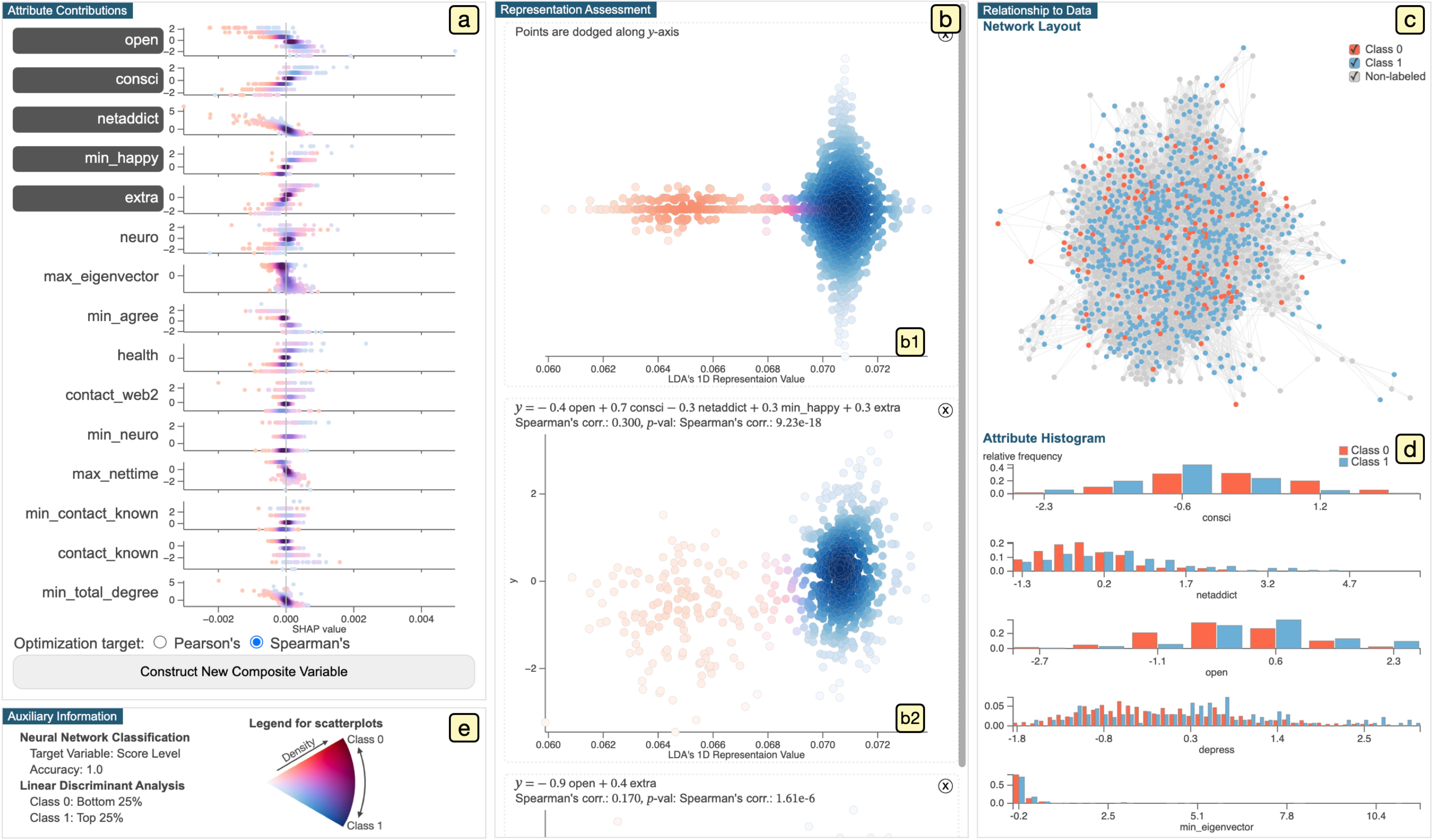}
    \caption{\CaptionUI{}}
    \label{fig:ui}
\end{figure*}

\subsection{Workflow Overview}

\noindent
\autoref{fig:workflow} shows our workflow for reviewing associations embedded in multivariate networks.
It consists of six steps that are first conducted with a script for ML (Steps 1--4) and then interactively performed with our user interface (UI) designed for visual analysis (Steps 5--6).
The UI is shown in \autoref{fig:ui}.
The UI presents the results from Steps 1--4 but is not used to perform interactive parameter adjustment for Steps 1--4.
These ML steps usually have various adjustable parameters (e.g., the numbers of NN layers and nodes) as well as require relatively long computation (e.g., 5 minutes in total). 
Consequently, for these steps, adjusting and running an ML script is more convenient to perform flexible analysis.

As we describe in \autoref{sec:repr_learning}, we extract structural information as a set of network measures (e.g., degree and betweenness centralities) for each element (i.e., node, link, or network); thus, in the following, we use a term, \textit{attribute}, to indicate both extracted structural (e.g., degree) and semantic information (e.g., age).

\textbf{Step 1.} 
The workflow begins with the extraction and selection of input attributes and one output/target attribute. 
For example, to know how students' grades in a school class are related to their surrounding conditions, an analyst can choose their attending class size, mental health status (e.g., depression level), and centrality in social media connections (e.g., degree) as inputs and their grade as an output.

\textbf{Step 2.} 
NRL is performed based on the selected inputs and output. 
A network representation is generated by an NN trained to perform binary prediction of the output (e.g., good or bad grade) with the inputs. 
The derived representation directly relates to associations of interest. 

\textbf{Step 3.} 
This step compresses the dimensionality of the network representation into one while preserving the prediction quality as much as possible. 
This makes the remaining steps simpler to complete and the related results easier to interpret.

\textbf{Step 4.} 
This step evaluates and ranks each input attribute's contribution to the 1D compressed representation. 
The obtained ranks can be considered as recommendation levels for the inclusion of the corresponding attributes for composite variable construction in Step 5. 

\textbf{Step 5.}
Finally, based on the recommendation levels and interests, an analyst manually selects a small set of attributes (e.g., 2--5 attributes) and runs a composite variable construction algorithm. 
The optimized composite variable (e.g., $y$-axis in \autoref{fig:ui}-b2) maximally resembles the 1D compressed representation (e.g., $x$-axis in \autoref{fig:ui}-b2). 
This composite variable provides an intuitive explanation of how the selected small set of attributes is related to the output/target attribute. 

\textbf{Step 6.}
The UI in \autoref{fig:ui} visualizes the information related to each previous step as well as the detailed structural and semantic information of the multivariate network. 
By interactively reviewing visualizations, the analyst can gain insights or adjust settings for each step based on their need (the backward arrows in \autoref{fig:workflow}).

\subsection{Two-class Density Scatterplot}
\label{sec:twoclass-scatterplot}

\begin{figure*}[t]
    \centering
    \includegraphics[width=0.92\linewidth]{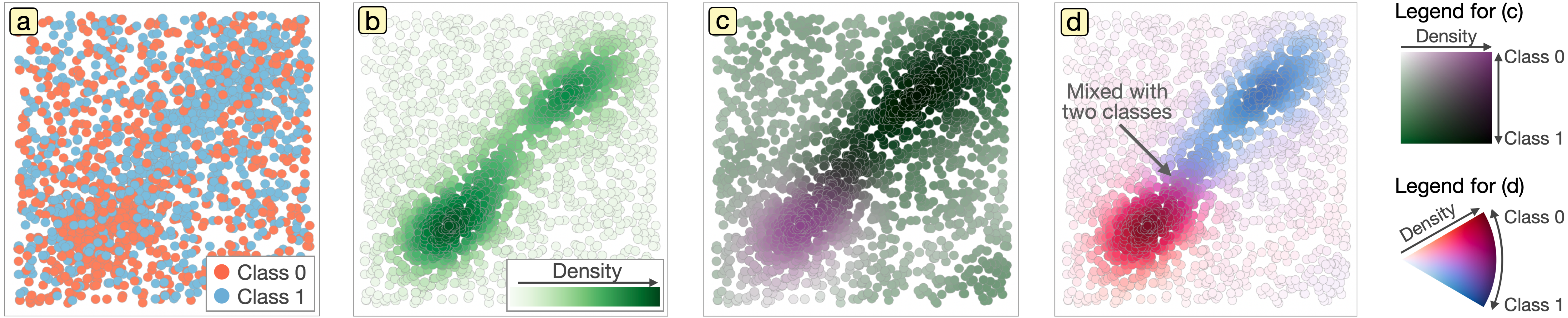}
    \caption{\CaptionScatterplots{}}
    \label{fig:scatterplots}
\end{figure*}

\noindent
We here introduce a \textit{two-class density scatterplot} (\autoref{fig:scatterplots}-d), which we use to visualize results from most of the steps in the workflow.
From a scatterplot of two variables (e.g., the 1D compressed representation and the composite variable generated at Steps 4--5), we aim to simultaneously depict (1) patterns related to binary classes; (2) trends including correlations; (3) other supplemental patterns, such as noises, outliers, and clusters.
For example, analysts may want to review class separation to assess the quality of network representations, correlation patterns related to the composite variable to gain insights, and patterns only seen in subgroups to consider further analyses.

According to a survey on scatterplot designs~\cite{sarikaya2018scatterplots}, we should employ an aggregated-level visual encoding (e.g., binning-based encoding) of instances to reveal trends while we should show individual instances for the other tasks (e.g., finding outliers).
However, designing an encoding that satisfies these requirements is not straightforward.
For example, in \autoref{fig:scatterplots}-a, class information is encoded with colored dots. 
This encoding easily suffers from overplotting when analyzing a large dataset (e.g., 1000 instances), hindering the observation of the data trend.
In contrast, density scatterplots are effective in finding trends.
For instance, \autoref{fig:scatterplots}-b reveals a correlated pattern.
However, from density scatterplots, we cannot grasp distributions related to classes (e.g., whether the dense area mostly consists of a single class).
A few existing scatterplot enhancements, splatterplots~\cite{mayprga2013splatterplots} and winglets~\cite{lu2020winglets}, can show both aggregated- and instance-level information along with class information. 
However, based on our experiment~\cite{supp}, the trend of data remains unclear in both enhancements.
Also, these enhancements are sensitive to the choice of parameters of their algorithms (e.g., threshold for contouring) and visualizations (e.g., dot size).

A two-class density scatterplot (\autoref{fig:scatterplots}-d) is designed to solve the above issues.
Using a bivariate colormap, a two-class density scatterplot encodes the \textit{total density} from both classes and the \textit{density ratio} of each class with color lightness and hue, respectively.
\autoref{fig:scatterplots}-d demonstrates how our design helps identify trends and class-related information from numerous instances.
First, it reveals the correlated pattern as in the density scatterplot.
In addition, we can confirm that two dense areas at the bottom left (more red colors) and top right (more blue colors) mainly consist of a single class.
Moreover, from the purple-color area annotated with the arrow, we can observe that a part of the dense areas mixes a considerable amount of both classes' instances.
We should note that a bivariate colormap~\cite{bernard2015colormap} has been already utilized to simultaneously show two different measures over scatterplots (e.g., \cite{lespinats2011checkviz}).
Our contribution is in the computation of the ratio of each class’s density and the color encoding design that is suitable to depict class and density information with scatterplots.

\textbf{Computation of density-related information.} 
Similar to an ordinary density scatterplot, we first estimate a 2D probability density function (PDF) from the coordinates of instances, using a Gaussian kernel density estimation. 
Let $\ProbDensFuncAll$ denote a 2D PDF estimated with all instances (i.e., the total density).
Then, the density at a 2D coordinate, $\Point$, can be computed with $\ProbDensFuncAll(\Point)$.
Similarly, let $\ProbDensFuncClassZero$ and $\ProbDensFuncClassOne$ denote 2D PDFs estimated with \texttt{Class\,0}'s instances and \texttt{Class\,1}'s instances, respectively. 
Also, let $\nInstsClassZero$ and $\nInstsClassOne$ be the numbers of instances in \texttt{Class\,0} and \texttt{Class\,1}.
Then, a function that estimates the density ratio of \texttt{Class\,0} at $\Point$ can be written as: $\DensRatioFuncClassZero(\Point) = \nInstsClassZero \ProbDensFuncClassZero(\Point) / (\nInstsClassZero \ProbDensFuncClassZero(\Point) + \nInstsClassOne \ProbDensFuncClassOne(\Point))$.
If assigning the same total density for each class is more appropriate for an analysis, we can use  $\DensRatioFuncClassZero(\Point) = \ProbDensFuncClassZero(\Point) / ( \ProbDensFuncClassZero(\Point) + \ProbDensFuncClassOne(\Point))$, instead.

\textbf{Visual encoding design.}
Using a bivariate colormap, we encode $\ProbDensFuncAll(\Point)$ and $\DensRatioFuncClassZero(\Point)$.
We select color lightness to encode the total density, $\ProbDensFuncAll(\Point)$, because of its natural metaphor: the more accumulation of dots, the darker color.
We then employ hue to represent the density ratio, $\DensRatioFuncClassZero(\Point)$.
To use lightness and hue, we initially used ordinary color spaces such as HSL; however, a subtle difference in the density was difficult to recognize from the generated colors. 
Thus, we decided to utilize existing carefully designed sequential colormaps.
Specifically, we use the red and blue sequential colormaps in Matplotlib~\cite{colamaps-in-matplotlib}.
We can obtain a pair of colors corresponding to  $\ProbDensFuncAll(\Point)$ from these two colormaps and then generate a linearly interpolated color from the pair based on $\DensRatioFuncClassZero(\Point)$.
To show this 2D color space as a legend, we use a polar coordinate (see \autoref{fig:scatterplots}).
Similar to Correl et al.’s colormap~\cite{correll2018valuesupressing}, this is to clearly indicate the difference between the meanings of two measures (total density and ratio) as well as to inform that the ratio difference is less emphasized in low total-density areas.
As shown in \autoref{fig:scatterplots}-c, we also tested existing bivariate colormaps such as one introduced by Lespinats and Aupetit~\cite{lespinats2011checkviz}.
This colormap was not suitable for two-density scatterplots. 
For example, it is difficult to recognize \texttt{Class\,1}'s high-density area and the density difference between the high-density areas of \texttt{Class\,0} and \texttt{Class\,1}.
Our supplementary materials~\cite{supp} provide more comprehensive comparison of various colormap designs.

\subsection{Datasets: Students and Adults’ Network Resources}
\label{sec:datasets}

\noindent
In the rest of this section, our explanations refer to a real-world dataset, Dataset I, as a concrete analysis target.
In addition to this dataset, we also analyze Dataset II in our case studies (\autoref{sec:case_studies}).
Both datasets are derived from the Facebook usage and survey data of representative cohorts in Taiwan.
While Dataset I is a single multivariate network, Dataset II consists of a set of egocentric networks.
The subjects authorized the use of their contact records on Facebook and survey answers. The use of data was approved by the IRB on Humanities \& Social Science Research, Academia Sinica, Taiwan.

\textbf{Dataset I} is about senior college students~\cite{chang2019social}.
We represented each student as a node and created a link based on their interactions on Facebook.
We constructed an undirected link if a ``like'', ``comment'', or ``tagged'' was made between students on a post. 
We consider such links as ``friendships'' in social media and 1-hop neighbors of each node as ``friends''.  
We further used the answers for the survey as each node's attributes.
The corresponding questions are related to the students' school life, such as their personalities, moods, and grades.
The resultant network consists of 1886 nodes, 64156 links, and 226 survey-based attributes for each node.

\textbf{Dataset II} is collected from sampled adults~\cite{lee2022indirect}.
This dataset uses Taiwan Social Change Survey, which includes important network-related topics, such as core discussion networks and network resources, as well as personal attributes, such as their backgrounds, personalities, moods, and social lives.
We generated multivariate egocentric networks by adding 1- and 2-hop neighbors of each survey respondent based on their contact records (i.e., 1-hop: direct contacts, 2-hop: indirect contacts).
We also assigned network-level attributes, such as the ego node's personality, Facebook usage statistics, and network measures (e.g., clustering coefficient).
The resultant egocentric networks consist of 345 ego nodes, 3339 1-hop neighbors, 19917 2-hop neighbors, and 11 network-level attributes.
In addition, we have information about whether each indirect contact is promoted to direct during a study period, January 1, 2015--June 30, 2017.

Below, from Dataset I, we identify and review college students' attributes that are highly related to their \texttt{scorelevel}---the past three years' mean score levels reported on a scale of four from the top to bottom. 
We used 14 survey attributes related to this analysis.

\subsection{Learning Representations}
\label{sec:repr_learning}

\noindent
We explain Steps 1--3, where we learn representations from multivariate networks. 
To provide a concise explanation, here we describe each process for producing a latent vector of each node (i.e., node feature learning).
However, the described learning processes can be easily adjusted to learn a latent vector for each link or network (\Flexibility).
\autoref{sec:cs3} demonstrates a case for network feature learning.

\subsubsection{Structural Feature Extraction for Attribute Selection}
\label{sec:preprocessing}

To select attributes related to the structural information in Step 1, we first precompute a set of structural measures~\cite{vandenelzen2016reducing,fujiwara2020visual}. 
For node feature learning, such measures can be degree, eigenvector, betweenness centralities, and many others~\cite{newman2018networks}.

These centralities inform the importance of each node from different aspects.
Degree centrality (i.e., the number of links connected to a node) measures local importance of nodes in the network.
Eigenvetor centrality (i.e., the eigenvector corresponding to the greatest eigenvalue of the network's adjacency matrix) more reflects global importance of nodes than degree centrality.
This is because eigenvector centrality considers the transitive importance of links (e.g., a link to a node with 100 links is more important than one with 1 link).
Betweenness centrality (the number of shortest paths that pass through a node) indicates a node's importance as a ``bridge'' connecting other nodes.
For details of the above three and other centralities (e.g., the Katz centrality and PageRank), refer to an introduction to network theory by Newman~\cite{newman2018networks}.

To capture more detailed structural information, we precompute statistics of each node's neighbors, such as the mean and variance of neighbors' degrees. 
Such statistics can be computed for semantic attributes as well (e.g., the mean age of neighbors).
The neighbor statistics are shown to be useful for various analytical tasks~\cite{rossi2018deep,fujiwara2020visual}.
For this precomputation, we specifically use DeepGL~\cite{rossi2018deep} to achieve fast computation as well as to avoid producing redundant measures. 

DeepGL summarizes 1-hop neighbors' centrality/attribute values (called \textit{base features}) with \textit{relational functions}.
For relational functions, we can specify multiple neighbor types (e.g., in-, out-, and total-neighbors) and aggregation functions (e.g., mean, variance, and maximum). 
After obtaining the neighbor statistics for all nodes, DeepGL prunes redundant statistics that show similar value distributions to others. 
The strength of pruning can be controlled with a hyperparameter of DeepGL.
Moreover, DeepGL can repeatedly apply the relational functions and pruning process to summarize farther neighbors' features. 
For more details, refer to the work by Rossi et al.~\cite{rossi2018deep}.
For Dataset I, as base features, we select 3 fundamental network centralities, degree, eigenvector, and betweenness, and the 14 survey attributes; then, we generate 1-hop total-neighbor statistics of the 17 attributes, resulting in 85 attributes in total after the pruning process.
Instead of the network centralities we selected, other centralities can be utilized based on a derived network and analysis purposes. 
For instance, when analyzing directed relationships, including the Katz centrality can be beneficial~\cite{newman2018networks}.

\subsubsection{Network Representation Learning}
\label{sec:nrl}

\noindent
After selecting input and output attributes in Step 1, we first apply the Z-score normalization to each attribute.
We then train an NN to predict the output from inputs, and extract representations from an NN layer (\Expressivity). 
Although we could predict an output value directly by designing the NN for a regression task, we instead categorize output values and perform binary classification of two value ends (e.g., top and bottom 25\% of attribute values). 
Since the classification is performed only on these ends, unrelated instances are not involved in training.
If we perform regression instead, the NN tries to fit all instances.
Consequently, a resulting representation may be only useful to predict values in the middle range (e.g., when fitting to the two ends is harder than the middle range).
While we assume this focus on two ends is reasonable for many analyses, NNs' settings are easy to adjust based on analytical interests.  
Another benefit of the binary-classification design is enabling a comparison of two categorical groups (e.g., students who major in engineering and business). 

By default, we use a multi-layer perceptron (MLP) consisting of five fully-connected layers (one input, three hidden, and one output layers) using the leaky rectified linear unit as an activation function. 
We then take the last hidden layer's activations as a learned representation.
With the consideration of a balance between expressivity and interpretability, we here keep the numbers and sizes of layers sufficiently small based on the size of the input data.
However, we should note that overfitting is acceptable for this step to some extent.
We simplify this step's high-dimensional network representation with linear transformation in Step 3 (\autoref{sec:repr_simplification}).
This simplification mitigates overfitting when it occurs.
The NN-based learning in Step 2 should focus on achieving an accurate alignment of the defined classes in a learned representation (i.e., high classification accuracy) rather than avoiding overfitting.
We discuss more details in \autoref{sec:repr_simplification}.

While the simple MLP is our default NN architecture, this architecture can be easily replaced with more complex ones (\Extensibility). 
For example, when analysis needs to capture nonlinear neighbor relations in a network, we can incorporate a GNN.
If learning latent vectors of links or networks is required, we can replace the precomputation of measures, accordingly.
For instance, we can extract the mean degree and network diameter for network feature learning.
Then, we can apply the same procedure employing an NN.

\subsubsection{Representation Simplification}
\label{sec:repr_simplification}

\noindent
In Step 3, we simplify the high-dimensional network representation into a 1D representation for the subsequent interpretation steps (\Interpretability).
Because the NN is trained for binary classification, this simplification is similar to the process NNs usually perform for their output layer: transforming the last hidden layer's activations of multiple NN nodes into a single NN node at the output layer.
Therefore, we expect that although the simplified representation is 1D, it can preserve sufficient information for the prediction (\Expressivity).

\begin{figure}[t]
    \centering
    \includegraphics[width=0.92\linewidth]{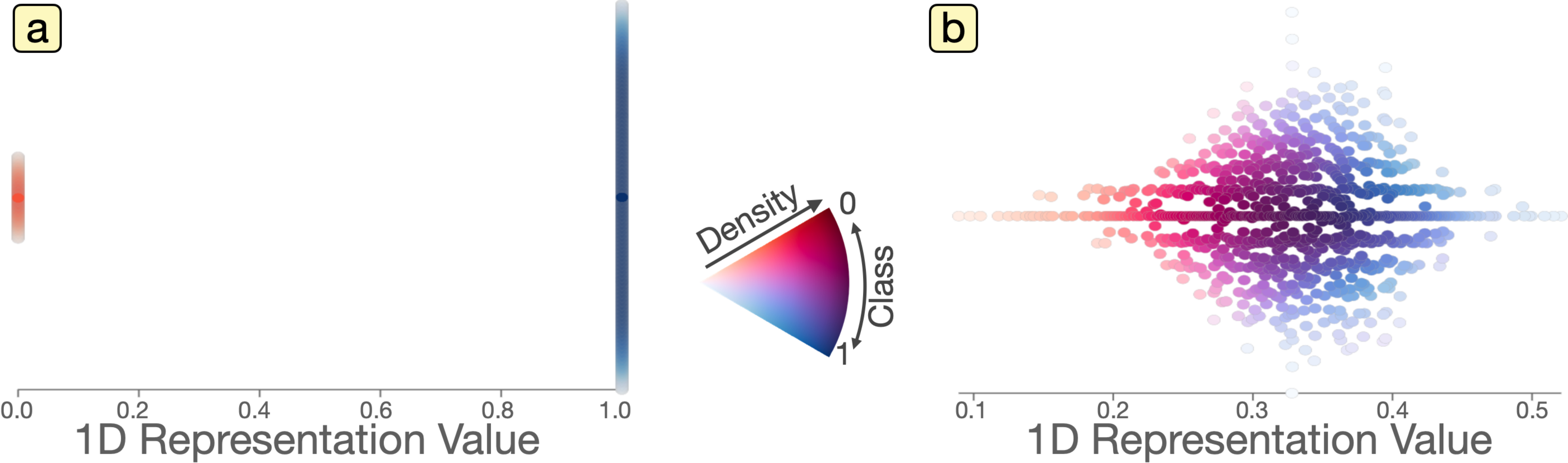}
    \caption{1D representations generated by alternative designs. The same dataset and visualization as \autoref{fig:ui}-b1 are used (red: \texttt{Class\,0}, blue: \texttt{Class\,1}). (a) is made by applying a softmax activation function at the last layer of an MLP. (b) is generated by directly applying LDA to the dataset.}
    \label{fig:alternatives}
\end{figure}

Our simplification employs linear transformation in contrast to nonlinear transformation.
This linear transformation is the major difference from conventional NNs. 
A nonlinear activation function such as a sigmoid or softmax function encourages producing a value of 0 or 1 for classification.
As a result, the corresponding 1D representations often consist of many values close to 0 or 1.
An example of such cases is shown in \autoref{fig:alternatives}-a.
This close-to-discrete distribution makes it difficult not only to perform optimization for composite variable construction but also to observe insightful patterns from the composite variable visualization shown in \autoref{fig:ui}-b. 

One way to obtain a 1D representation with linear transformation is using a linear activation function between the last hidden and output layers, as in NNs for regression tasks.
As stated in \autoref{sec:nrl}, we, however,  want to avoid directly performing regression in the NN. 
Instead, as a post-hoc process, we apply a DR method to the high-dimensional network representation.
We specifically use LDA, which places instances of different classes as far as possible in a low-dimensional representation. 
This placement is achieved by maximizing the distance of each class’s centroid while minimizing the variance of each class.
This post-hoc process can efficiently generate a 1D representation with optimal class separation.

On the other hand, even though feature selection methods, such as LDA, are low in computational complexity, LDA can still cause overfitting when the number of instances is relatively small to the number of input features~\cite{guo2007regularized}. 
Thus, we further utilize regularized LDA, which has a regularization mechanism to avoid overfitting~\cite{guo2007regularized}.
The use of regularized LDA brings an additional benefit for the post-hoc approach: the regularization strength can be controlled outside of the NN training process. 
Although the $L1$ and $L2$ regularizations can also be used for NNs to avoid overfitting~\cite{zou2005regularization}, the hyperparameter adjustment for these regularizations often requires many trials and errors, and becomes time-consuming by retraining NNs many times.
Applying regularization inside of LDA does not require retraining of NNs and LDA is computationally efficient; thus, we can adjust the regularization strength more conveniently (\Steerability).

The quality of the 1D representation can be investigated with a swarm-plot-like visualization, as shown in \mbox{\autoref{fig:ui}-b1}.
This visualization shows an instance as a dot and a 1D representation value as an $x$-coordinate.
Similar to the swarm plot~\cite{swarmplot}, $y$-direction is used to pile up dots with positional jitters to mitigate overplotting with a limited vertical space.
The resultant area height roughly shows the frequency/density around the corresponding $x$-coordinate (similar to a histogram).
We select this instance/point-based visualization to keep it consistent with other visualizations that use $x$-coordinates to represent the 1D representation values (cf. \autoref{fig:ui}-b2).
The color encoding is as with two-class density scatterplots'.
When the 1D representation has good quality, as seen in \autoref{fig:ui}-b1, two classes (\texttt{Class\,0}: red, \texttt{Class\,1}: blue) should have small or no overlaps. 
When the quality is not satisfactory, the analyst can update the NN architecture and/or the selection of input and output attributes.
For example, the analyst should consider increasing the number of NN nodes and/or layers when the data has many instances and attributes.
Similarly, if there are more attributes that NNs can utilize, the inclusion of them might reduce the overlaps.

\begin{figure}[t]
    \centering
    \includegraphics[width=0.9\linewidth]{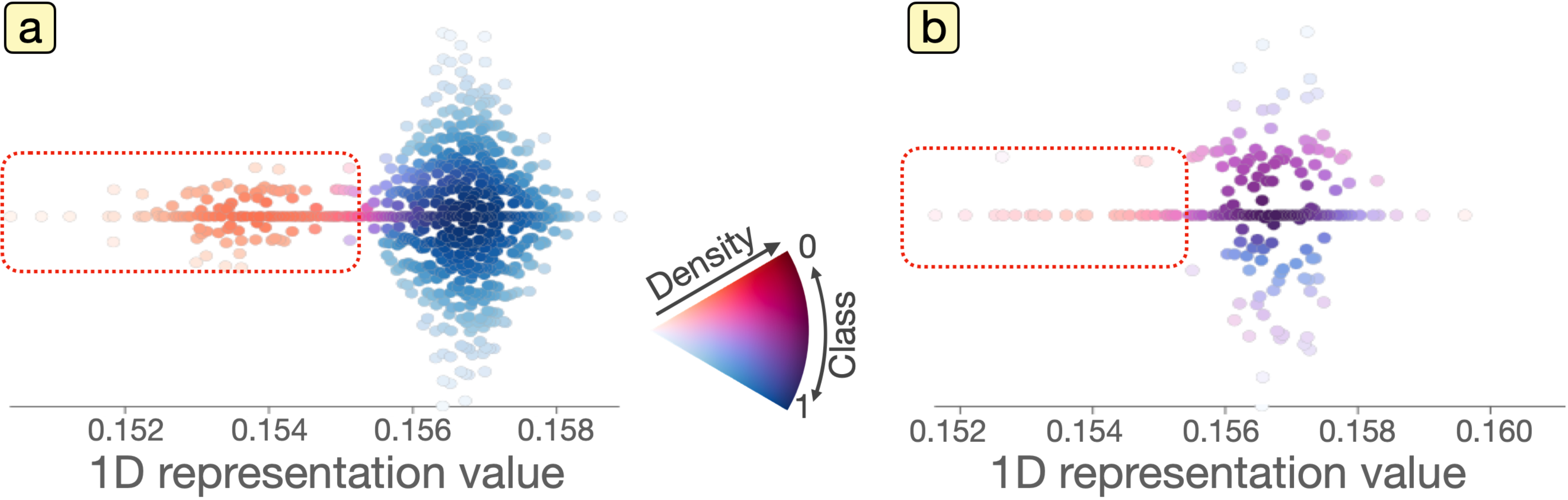}
    \caption{1D representations generated by training the same NRL architecture as the \autoref{fig:ui}-b1. 
    Unlike \autoref{fig:ui}-b1, the NRL architecture was trained with a subset of the data (616 instances). 
    (a) and (b) correspond to the 1D representations of the training set and the remaining test set (170 instances).}
    \label{fig:overfit_train_test}
\end{figure}

The 1D representation in \autoref{fig:ui}-b1 exhibits how the simplification using linear transformation relaxes the overfitted representation.
Unlike the representation directly output from the MLP shown in \autoref{fig:alternatives}-a, \autoref{fig:ui}-b1 shows a continuous distribution of \texttt{Classes\,0} and \texttt{1}.
To further check if a simplified representation captures general data trends, we experimentally generated 1D representations while using separate training and test datasets.
The resultant 1D representations are shown in \autoref{fig:overfit_train_test}.
In \autoref{fig:overfit_train_test}-a, we can see the 1D representation generated from only a subset of the data (i.e., training set) still shows a similar pattern to \autoref{fig:ui}-b1.
Then, in \autoref{fig:overfit_train_test}-b, although there is a less clear separation between \texttt{Classes\,0} and \texttt{1}, we can still observe a similar pattern to \autoref{fig:overfit_train_test}-a. 
For example, a considerable amount of instances of \texttt{Class\,0} is placed on the left side.
The above observations show that the potential issue of overfitting in MLP becomes less problematic through representation simplification.

In addition to the discussed architectures, we experimented with another alternative design: directly applying LDA to data to generate a 1D representation.
As shown in \autoref{fig:alternatives}-b, because this design did not exploit the strengths of NNs, the results often mixed two classes.

\subsection{Understanding Representations}
\label{sec:repr_interpretation}

\noindent
We describe Steps 4 and 5, which are mainly for understanding the representation obtained through Steps 1--3 (\Interpretability).

\subsubsection{Attribute Contribution Measurement}
\label{sec:attrib_contrib_shap}

\noindent
In Step 4, we extract each input attribute's contribution to the 1D representation, which is useful for two tasks.
First, this attribute contribution is useful to understand each attribute's relationship to the output attribute---the interpretation from a \textit{single attribute} level. 
Second, the contribution indicates attributes that should be considered for composite variables---the interpretation considering influence from \textit{multiple attributes}.

We employ the SHAP method~\cite{lundberg2017shap} to measure the attribute contributions.
The SHAP method calculates a measure, called the SHAP value for each instance.
The SHAP value is computed based on cooperative game theory and shows how much a target value will be changed by adding each attribute value in a model (refer to \cite{lundberg2017shap} for details).
We apply the SHAP method to a transferred NN model that combines the MLP trained in Step 2 and LDA trained in Step 3.
In this case, the SHAP value indicates how much having a corresponding attribute value contributes to moving an instance toward a positive direction of the 1D representation.
For example, when an instance's \texttt{age} is 25 and its SHAP value for \texttt{age} is 0.1, having this age shifts the instance to a more positive side of the 1D representation by the magnitude of 0.1.

The SHAP method is suitable for our workflow for two reasons. 
First, the SHAP's model-agnostic property enables extracting the attribute contributions from our custom model combining the NN and LDA.
Second, compared with another model-agnostic method, LIME~\cite{ribeiro2016should}, the SHAP method generates attribute contributions that can be compared across instances.
This analytical capability is important to understand the general influence of each attribute on the output (e.g., internet addiction's negative influence on score levels).

\begin{figure}[t]
    \centering
    \includegraphics[width=\linewidth]{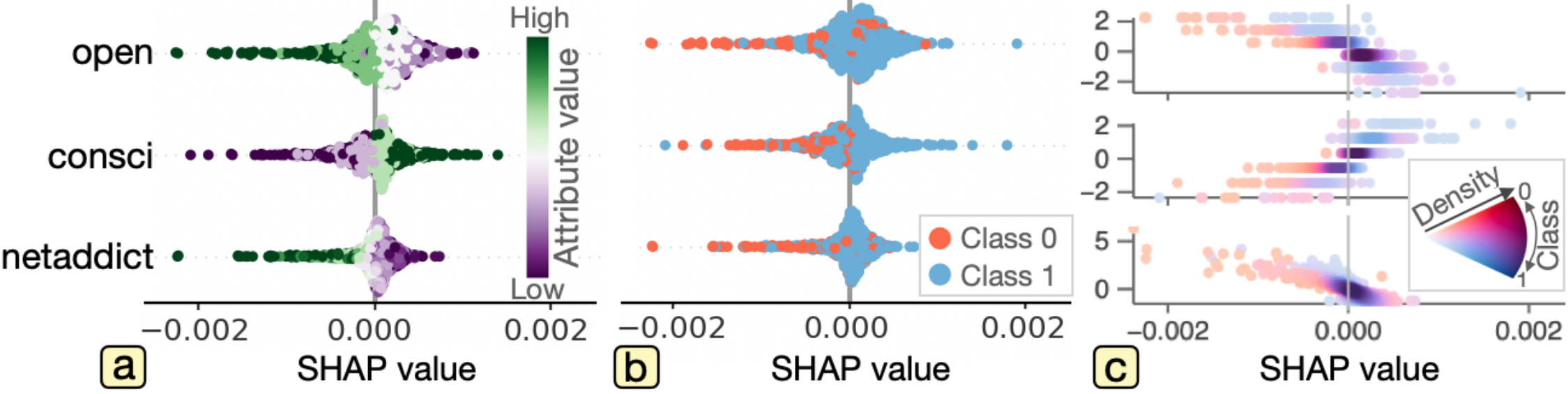}
    \caption{\CaptionSHAP{}}
    \label{fig:shap}
\end{figure}

\textbf{Visualization.}
As shown in \autoref{fig:ui}-a (also \autoref{fig:shap}-c), we visualize SHAP values with a list of two-class density scatterplots, where each row corresponds to one attribute's SHAP values.
Our design uses $x$- and $y$-axes to depict the SHAP values and attribute values, respectively.
As a summary measure of each attribute's contribution, we use the mean absolute value of all instances' SHAP values.
We then sort the display order of the scatterplots based on this summary measure (i.e., top contributed attributes are listed from the top in order).

For example, in \autoref{fig:ui}-a, we see the top 3 contributed attributes are \texttt{open} (openness), \texttt{consci} (conscientiousness), and \texttt{netaddict} (internet addiction). 
Furthermore, from the corresponding scatterplots, we can gain several insights. 
While larger \texttt{open} has a more negative impact on the 1D representation reflecting \texttt{scorelevel}, larger \texttt{consci}  has a more positive impact.
Also, \texttt{netaddict} shows clearly different patterns by \texttt{Class\,0} (low \texttt{scorelevel}) and \texttt{Class\,1} (high \texttt{scorelevel}).
Larger \texttt{netaddict} tends to have a more negative impact for \texttt{Class\,0}. 

We tested different visual designs, as shown in \autoref{fig:shap}.
\autoref{fig:shap}-a follows the default plot in the SHAP Python package~\cite{shap_library}, where the swarm-plot-like visualization is used. 
$x$-coordinates correspond to SHAP values, while colors inform attribute values.
From this visualization, we can roughly see the relationships between these values (e.g., larger \texttt{open} tends to have a more negative SHAP value).
But, it is infeasible to grasp patterns that differ by class.
Even when juxtaposing a visualization that color-encodes class information, as shown in \autoref{fig:shap}-b, we still cannot clearly see such different patterns. 
In contrast, our final design (\autoref{fig:shap}-c) can clearly depict the relationships between SHAP and attribute values as well as different patterns by class.

\subsubsection{Composite Variable Construction}
\label{sec:comp_var_construction}

\noindent
In Step 5, we construct composite variables interactively.
The composite variable is commonly used when using a single attribute is not sufficient to understand a target phenomenon ({\Expressivity} and {\Interpretability}). 
For example, in baseball, on-base plus slugging (OPS)---the sum of two different statistics of the player's performance, on-base percentage and slugging percentage---is used to analyze players' contributions to team runs.
This is because OPS is a simple composite variable but extremely correlates to team runs~\cite{albert2010sabermetrics}. 

\textbf{Construction algorithm.} We describe the optimization algorithm designed to construct a composite variable from user-selected attributes.
To explain the 1D representation with the selected attributes as much as possible, our optimization goal is to maximize a certain \textit{measure of dependence}~\cite{reshef2016measuring,li2020multimodel} between the 1D representation and attributes. 
Such measures include Pearson's correlation, Spearman's correlation, and mutual information. 
We constrain a composite variable to a linear combination of attributes to achieve a simple, intuitive interpretation as well as an efficient optimization for interactive construction.

A linear combination that achieves the highest Pearson's correlation coefficient can be derived by multivariate linear regression. 
On the other hand, Spearman's involves a non-differentiable operation to rank the 1D representation values.
Due to non-differentiability, we cannot use simple regression or gradient-based solvers.
Thus, we employ a gradient-free solver, specifically COBYLA (or constrained optimization by linear approximations)~\cite{powell1998direct} while using the optimized result for Pearson's as an initial solution. 
Although the optimization for Pearson's is much more efficient, Spearman's could be more effective in some cases.
This is because the 1D representation is constructed from the non-linearly transformed attributes by the NN, and the 1D representation may have a nonlinear correlation with the original input attributes.
Although we currently only support Pearson's and Spearman's correlations, the algorithm can be easily extended for other measures of dependence (e.g., mutual information) by using COBYLA or a gradient-based solver.

\textbf{Visualization and interactive use.} From the view in \mbox{\autoref{fig:ui}-a}, an analyst can select attributes and a measure of dependence, and then generate a composite variable by clicking ``Construct New Composite Variable'' located at the bottom.
The selected attributes are highlighted with gray backgrounds.
Then, a two-class density scatterplot (\autoref{fig:ui}-b2) visualizes the relationships between the 1D representation ($x$-axis) and the constructed composite variable ($y$-axis).
Also, multiple composite variables can be created to compare their expressivity and interpretability (\Steerability). 
A newly constructed composite variable's visualization is appended to a list of the scatterplots in a scrollable view.
Each scatterplot can be discarded by clicking a ``$\smash{\times}$'' mark at the top right.  

In essence, through this composite variable construction process, analysts can utilize their domain knowledge to review the simplified representation.
From the attributes ranked by SHAP values, the analyst can judge attributes they should consider.  
Then, with the help of the above optimization algorithm, the analyst can interactively examine and fine-tune the composite variables to gain insights.
Our contribution to the interpretation of representations lies in this analysis flow, starting from the SHAP ranking of individual attributes, incorporating analysts' judgment, and performing the construction and interpretation of composite variables.

\textbf{Analysis examples.}
For Dataset I, we compute Spearman's correlation coefficient between each of the top 5 contributed attributes and the 1D representation. The cofficients are \texttt{open}:\,\mbox{-0.147}, \texttt{consci}:\,0.237, \texttt{netaddict}:\,0.134, \texttt{min\_happy} (the minimum of friends' happiness levels):\,0.093, and \texttt{extra} (extraversion):\,0.026.
As shown in \autoref{fig:ui}-b2, using all these attributes, we can generate a composite variable that has a better correlation coefficient, 0.300, than any single attribute.
This coefficient is in a moderate correlation group based on Dancey and Reidy's categorization~\cite{dancey2017statistics}.
We follow Dancey and Reidy's categorization in the rest of the paper when describing correlation strengths.

Since all attributes are normalized, each attribute's weight in the composite variable shows a contribution level to the composite variable (\texttt{open}:\,-0.4, \texttt{consci}:\,+0.7, \texttt{netaddict}:\,\mbox{-0.3}, \texttt{min\_happy}:\,+0.3, \texttt{extra}:\,+0.3).
\texttt{consci} contributes most, as expected from its higher Spearman's correlation (0.237) than others. 
In contrast, \texttt{extra} is assigned a relatively large weight (+0.3) in spite of its extremely small correlation (0.026) as a single attribute.
The importance of \texttt{extra} can be confirmed by excluding \texttt{extra} from the composite variable construction. 
When \texttt{extra} is extruded, the composite variable's correlation coefficient decreases from 0.300 to 0.289. 
From further interactive investigations on how the inclusion of \texttt{extra} improves each other attribute's correlation, we observe that \texttt{open} has a much larger improvement (from 0.147 to 0.170) than others (e.g., \texttt{consci} has no improvement).
We can infer that, by including \texttt{open} and \texttt{extra} with different signs (-0.4 and 0.3), the composite variable captures the subtle difference between these two personalities.
Then, \texttt{scorelevel} shows higher dependence on this derived difference than either \texttt{open} or \texttt{extra}. 
This type of insight cannot be derived if only investigating a single attribute relationship to the 1D representation. 
Also, this example demonstrates the effectiveness of the SHAP-based recommendation in contrast to relying only on each attribute's correlation coefficient.

We should emphasize that while we want to make a composite variable correlated to the 1D representation to some extent (e.g., Spearman's $\smash{\geq 0.2}$, weakly correlated), we do not have to see extremely correlated results (e.g., Spearman's $\smash{\geq 0.7}$, strongly correlated).
If a strong correlation can be found with a linear combination of a few attributes, the use of NNs becomes unnecessary.
In such a case, linear learning methods such as LDA should be sufficient for the prediction. 
Our focus is taking a further step from the single-attribute-based interpretation using the SHAP or other methods~\cite{lundberg2017shap,kwon2018clustervision,neto2021multivariate}.
Our approach considers the influence of multiple attributes. 
Such an influence can be analyzed via the signed weights in a composite variable and the interactive adjustment of the composite variable. 
The correlations to each composite variable should be considered as an indicator of how much of the 1D representation is explained with the composite variable.

\subsection{Interactive Visual Interface}

\noindent
We have introduced each visual component with its corresponding workflow step in \autoref{sec:repr_learning} and \autoref{sec:repr_interpretation}.
In summary, the selected output (Step 1) and performance of NRL (Step 2) are shown as the auxiliary information in \mbox{\autoref{fig:ui}-f}; and the simplified representation (Step 3) is visualized in \autoref{fig:ui}-b1. 
Then, to aid in understanding the representation, \autoref{fig:ui}-a depicts the attribute contribution information (Step 4) and serves as an interface to construct composite variables (Step 5). 
The created composite variables are visualized in \autoref{fig:ui}-b2.
As we explain in this section, the remaining views in \autoref{fig:ui}-c and d depict structural and semantic information to supplement the interpretation of the results.
All visualizations in the UI are fully linked and share the same or similar color encodings (e.g., red represents \texttt{Class\,0}). 
Lasso selection can be performed in \autoref{fig:ui}-a, b, and c. 
The selected instances from \texttt{Classes\,0} and \texttt{1} will be highlighted in yellow (e.g., \mbox{\autoref{fig:case2}-c}).
With this linking, analysts can investigate specific patterns, such as outliers and subgroups.

\textbf{Structural information.} 
\autoref{fig:ui}-c shows network layouts as the structural information. 
Only in this view, we visualize all instances even including those not belonging to the two ends (i.e., \texttt{Classes\,0} and \texttt{1}).
This is because the precomputed network measures (e.g., degree) used for NRL are computed using links among all instances.
And, we need to review the entire network to locate patterns related to the structural information.
We apply scalable force-directed placement (SFDP)~\cite{hu2005efficient} and then use red, blue, and gray colors to represent \texttt{Class\,0}, \texttt{Class\,1}, and other instances, respectively.
For cases when analysts want to review relationships in a specific class, the UI also supports filtering of instances.
The filtering can be applied by using colored checkboxes located at the top right.

\textbf{Attribute information.} 
We visualize the distribution of each attribute as a double-bar histogram. 
By default, as shown in \autoref{fig:ui}-d, the UI shows the distribution for each class.
When the lasso selection is performed, we show the distributions of selected and non-selected instances (e.g., \autoref{fig:case1}-b1 and b2).
Because we have limited screen space, we order histograms based on the two groups' distribution differences measured by the Kolmogorov-Smirnov (KS) statistic.

\textbf{Implementation.}
The UI is developed as a web application. 
For the back end, Python is used to perform Step 1 with DeepGL, Step~2 with MLPs, Step 3 with regularized LDA, Step 4 with the SHAP method, Step 5 with multivariate regression and the COBYLA, and Step 6 with the two-class density scatterplot, SFDP, and KS statistic.
The implementation of these algorithms utilizes various libraries, such as deepgl~\cite{fujiwara2022network}, PyTorch~\cite{pytorch2019paszke} (for MLPs), ulca~\cite{fujiwara2021interactive} (for regularized LDA), SHAP~\cite{shap_library}, Scikit-learn~\cite{pedregosa2011scikit} (for multivariate regression), NumPy/SciPy~\cite{virtanen2020scipy} (for the COBYA, KS statistic, and Gaussian kernel density estimation), graph-tool~\cite{graphtool2014peixoto} (for SFDP), seaborn~\cite{waskom2021seaborn} (for the swarm-like visualization), and ColorAide~\cite{coloraide} (for color generation).
The front-end UI is implemented with HTML5, JavaScript, and D3~\cite{bostock2011d3}. 
WebSocket is used to communicate the front- and back-end modules.

\section{Case Studies}
\label{sec:case_studies}

\noindent
We demonstrate the effectiveness of our workflow and interactive visualizations with three case studies using the two datasets.
For the first two case studies on Dataset I, we show analyses on network nodes.
In the third case, we analyze the egocentric networks of Dataset II.

\subsection{Study 1: Associations with Score Levels}
\label{sec:cs1}

\noindent
We present a complete version of the analysis we have performed in \autoref{sec:repr_learning} and \autoref{sec:repr_interpretation}. Our analysis target is the identification of attributes that are highly related to college students' \texttt{scorelevel}.
We classify the bottom and top \texttt{scorelevel} groups with the five-layer MLP using 128, 64, and 64 NN nodes for hidden layers.
138 and 648 students are categorized into the bottom group (\texttt{Class\,0}, red color) and top group (\texttt{Class\,1}, blue color), respectively.
The prediction results show 1.0 accuracy for Step 2 but 0.98 for Step 3 (refer to \autoref{sec:nrl} for the reason why we accept such high accuracies). 

After completing Steps 1--3, we first confirm that the extracted 1D representation provides a reasonable separation between the two classes, as shown in \autoref{fig:ui}-b1.
Also, the distribution of instances/students shows a sufficient variety in their coordinates. 
Thus, we expect that the 1D representation is not extremely overfitted for this classification.

We then review attributes' associations with the 1D representation from the view in \autoref{fig:ui}-a.
Based on the ranked order of the attributes, contributing attributes to the score level differences are highly related to many of the Big Five personality traits: \texttt{open} (openness), \texttt{consci} (conscientiousness), \texttt{extra} (extraversion), \texttt{neuro} (neuroticism), and \texttt{agree} (agreeableness).
Other highly ranked attributes include
\texttt{netaddict} (internet addiction level) and \texttt{min\_happy} (the minimum of the friends' happiness levels).
On the other hand, most network centralities are not ranked high. 
As we examined in \autoref{sec:attrib_contrib_shap}, the top 5 attributes have clear positive or negative trends with the SHAP values (e.g., higher \texttt{open} value tends to have a more negative SHAP value).

Next, we construct a composite variable with the top 5 attributes while using Spearman's as a measure of dependence.
\autoref{fig:ui}-b2 visualizes the result.
The composite variable is expressed as: $y$ = -\,0.4\,\texttt{open} +\,0.7\,\texttt{consci} -\,0.3\,\texttt{netaddict} +\,0.3\,\texttt{min\_happy} +\,0.3\,\texttt{extra}.
Each weight's sign agrees with the corresponding attribute's positive or negative influence on the 1D representation, as observed in \autoref{fig:ui}-a.
Based on the magnitudes, we can see \texttt{consci} contributes most to the 1D representation.
Also, as discussed in \autoref{sec:comp_var_construction}, we observe that \texttt{open} and \texttt{extra}---personalities that might have some overlapped aspect---likely derive a new meaningful attribute by having opposite signs with each other.
When a composite variable is constructed with these two attributes, the resultant variable, -\,0.9\,\texttt{open} +\,0.4\,\texttt{extra}, shows a 0.17 correlation coefficient. 
Thus, having both openness and introversion tends to negatively impact their score levels.

\begin{figure}[tb]
    \centering
    \includegraphics[width=\linewidth,height=0.53\linewidth]{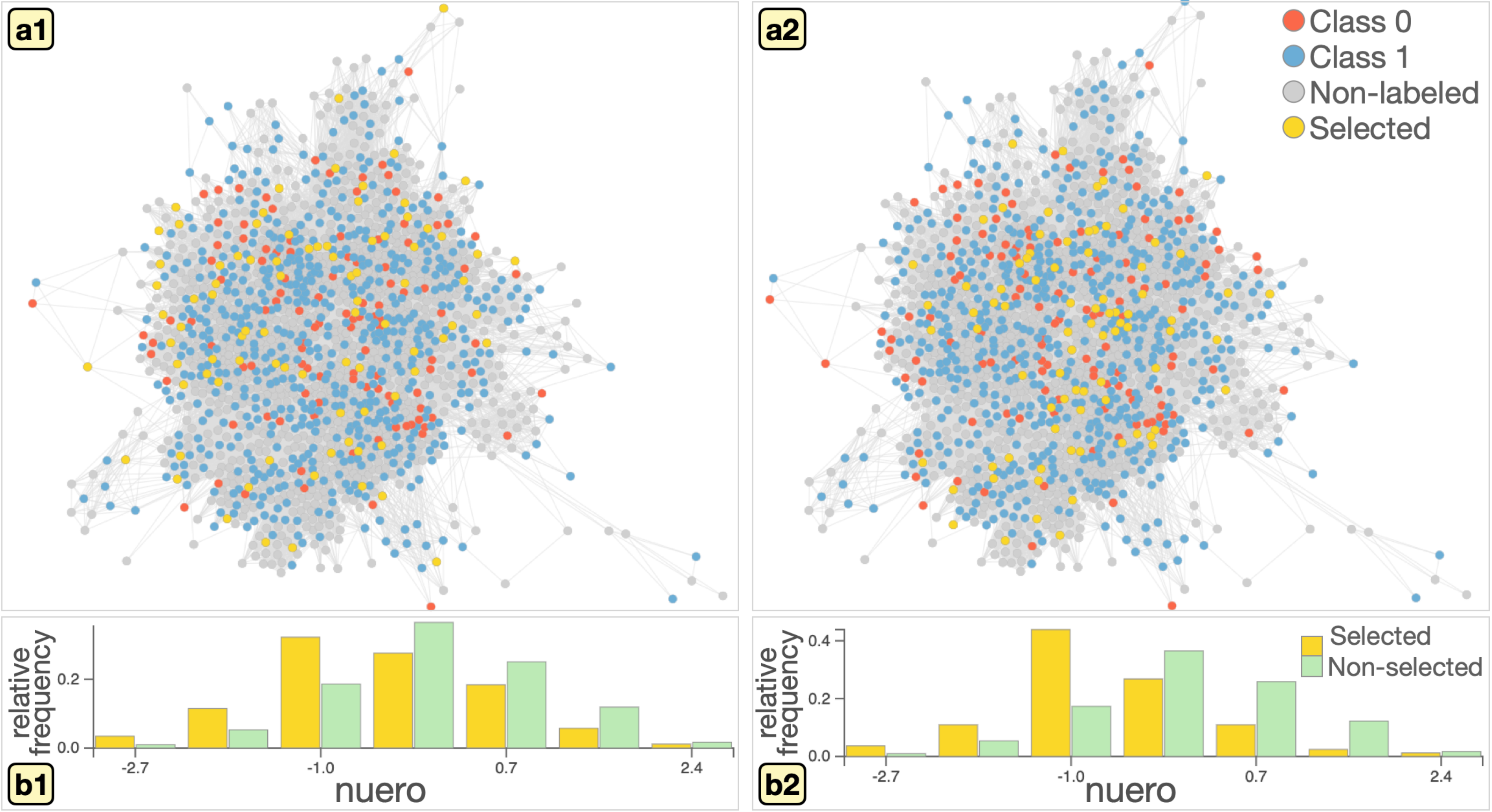}
    \caption{\CaptionCaseOne{}}
    \label{fig:case1}
\end{figure}

We investigate another question raised during the above analysis: why network centralities were not listed as highly contributed attributes even though
the composite variable, \mbox{-\,0.9}\,\texttt{open} +\,0.4\,\texttt{extra}, shows a clear contribution to the 1D representation.
Openness and extraversion could be related to network centralities such as node degree (e.g., extrovert students are likely to have more friends).
To examine the relationships to network structure, we highlight students with low (-\,0.9\,\texttt{open} +\,0.4\,\texttt{extra}) in the network layout visualization, as shown in \autoref{fig:case1}-a1.
While we see several selected students (colored yellow) are located on the outskirts of the network (i.e., fewer connections to others), we cannot find any clear structural pattern. 
When selecting students with high \texttt{extra} (\autoref{fig:case1}-a2), we cannot see a clear pattern either.
Thus, we can expect that openness and extraversion in real life were difficult to capture only from the connections on Facebook.  

\begin{figure*}[tb]
    \centering
    \includegraphics[width=\linewidth]{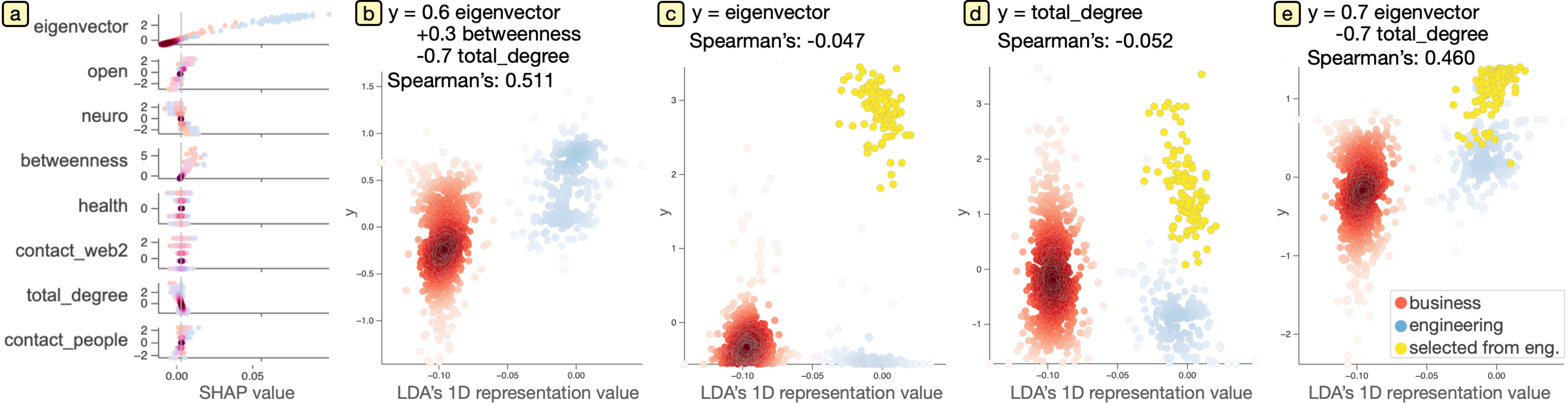}
    \caption{\CaptionCaseTwo{}}
    \label{fig:case2}
\end{figure*}

\begin{figure}[tb]
    \centering
    \includegraphics[width=\linewidth]{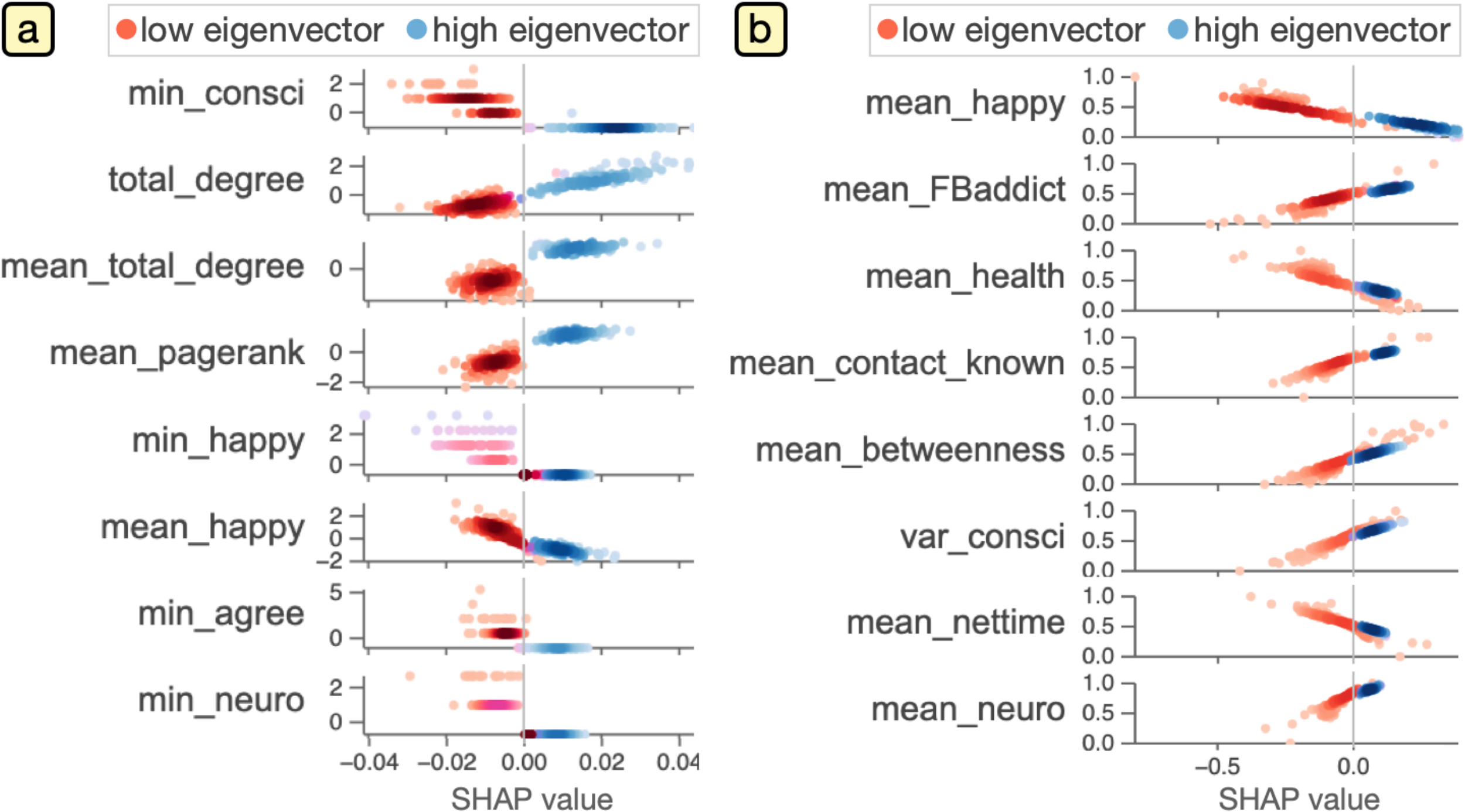}
    \caption{\CaptionCaseTwoCont{}}
    \label{fig:case2_2}
\end{figure}

We further see the UI suggests \texttt{neuro} shows strongly different distributions between the selected and
non-selected students.
From the distributions of \texttt{neuro} shown in \mbox{\autoref{fig:case1}-b1} and b2, the students with high (-\,0.9\,\texttt{open} +\,0.4\,\texttt{extra}) or high $\texttt{extra}$ tend to have lower \texttt{neuro} than others. 
Therefore, \texttt{neuro} is also interrelated to \texttt{open} and \texttt{extra}.
However, based on \autoref{fig:ui}-a, \texttt{neuro} has a smaller influence on the score level and does not show consistent positive or negative influences among students.
In fact, the inclusion of \texttt{neuro} into the composite variable of the top 5 attributes does not improve the correlation coefficient.

The above observations derive several reasonable insights: The student's conscientiousness is highly related to their score level; internet addiction has a negative association with the score, especially, for those who had bad scores (refer to \autoref{sec:attrib_contrib_shap}); if all friends have sufficient happiness, the score tends to be higher, and vice versa; and the openness and extraversion show a clear combinational effect and a high openness with a low extraversion has a more negative relationship to the score.

\subsection{Study 2: Differences in Academic Units}

\noindent
From Dataset I, we review whether students from different majors have different structural and semantic characteristics.
To perform this comparison, we select each pair of all possible different majors (\texttt{agricultural}, \texttt{business}, \texttt{engineering}, \texttt{humanities}, \texttt{physical\,sciences}, and \texttt{social\,sciences}). 
We reviewed all pairs, but here we show one representative analysis: the comparison of \texttt{business} (\texttt{Class\,0}, red, 801 students) and \texttt{engineering} (\texttt{Class\,1}, blue, 287 students).
We use the same 85 attributes and MLP as Study~1 and perform classification of \texttt{business} and \texttt{engineering}.
The prediction results show 1.0 accuracy for both Steps 2--3. 

\autoref{fig:case2}-a shows the top 8 contributed attributes to the differences between \texttt{business} and \texttt{engineering}.
Unlike Study 1, this list of the top 8 includes network centralities, \texttt{eigenvector}, \texttt{betweenness}, and \texttt{total\_degree}.
To see the associations between these centralities and the 1D representation, we generate a composite variable with these three centralities, as shown in \autoref{fig:case2}-b.
The resultant composite variable shows a moderate correlation (Spearman's: 0.511). 
Also, we can see large weights with opposite signs for \texttt{eigenvector} and \texttt{total\_degree} (+0.6 and -0.7).
Unlike degree, eigenvector centrality considers the importance of links (discussed in \autoref{sec:preprocessing}). 
Taking a subtraction of degree from eigenvector centrality can emphasize this unique characteristic of eigenvector centrality.

As \autoref{fig:case2}-a depicts a clear positive influence of high \texttt{eigenvector}, we visualize the relationships between \texttt{eigenvector} and the 1D representation.
From the result shown in \autoref{fig:case2}-c, we first notice that \texttt{eigenvector} itself has a very small correlation coefficient (\mbox{-0.047}). 
At the same time, there are two clear subgroups in a group of \texttt{engineering} as we highlight one group in yellow.
On the other hand, as shown in \autoref{fig:case2}-d, a similar but more moderate separation can be seen in \texttt{total\_degree}. 
We then construct a new composite variable with only \texttt{eigenvector} and \texttt{total\_degree}.
The result shown in \autoref{fig:case2}-e informs a clear correlation of this new composite variable to the 1D representation (+0.460).
From the visualizations in \autoref{fig:case2}-c, d, and e, we now grasp how this informative composite variable for the difference of \texttt{business} and \texttt{engineering} is generated from the two centralities. 
This result emphasizes the importance of interpretation using a combination of multiple attributes. 
It also highlights the benefits of the design of a two-class density scatterplot, which aids in identifying denser clusters and inferring trends and correlations.

We are further interested in understanding the difference between low and high \texttt{eigenvector} groups in \texttt{engineering} students.
To perform this analysis, we select these two groups as a classification target.
\autoref{fig:case2_2}-a shows the derived top 8 contributing attributes, where red and blue colors are now used for low and high \texttt{eigenvector} groups. 
Most attributes show a clear separation between the groups.
Also, many of them are related to the student's personality or mentalities, such as \texttt{consci}, \texttt{happy}, and \texttt{agree}. 
Meanwhile, the listed attributes in \autoref{fig:case2_2}-a tend to correspond with the minimum values of their friends', as indicated with the prefix, ``\texttt{min\_}''.
As the high \texttt{eigenvector} group tends to have high \texttt{total\_degree} (see \autoref{fig:case2}-d), the students in the high \texttt{eigenvector} group are likely to have many friends, resulting in a higher chance to have at least one friend with low values for these attributes.
Therefore, we consider that this result is caused by inappropriate fittings specific to these two groups.
Also, as already observed in \autoref{fig:case2}-d, the two groups have clearly different values in \texttt{total\_degree}.
As the inclusion of these attributes is problematic, we remove them during the feature extraction and rerun the remaining steps.
\autoref{fig:case2_2}-b shows the resultant top 8 contributing attributes.
By reviewing the values of each attribute (i.e., $y$-coordinates in \autoref{fig:case2_2}-b), we infer that friends of the high \texttt{eigenvector} group tend to have lower happiness levels (\texttt{happy}), worse health status (\texttt{health}), and are more addicted to Facebook (\texttt{FBaddict}), while they spend shorter time on the Internet (\texttt{nettime}).
These results highlight the importance of considering compounded factors in network analysis (e.g., more social media friends but the friends are less happy; more addicted to social media but shorter time spent on the Internet). 

From this study, we observe the difference between \texttt{business} and \texttt{engineering} students in the composite variable constructed with the network centralities. 
As the composite variable emphasizes the difference between the eigenvector and total degree centralities, \texttt{engineering} students' connections tend to be more particular to influential students in Facebook. 
Also, we demonstrate the case of identifying and resolving potential inappropriate overfitting from the visualized results.

\subsection{Study 3: Factors Promoting New Connections}
\label{sec:cs3}

\noindent
From Dataset II containing 345 egocentric networks of adults, we review contributing factors to promoting their indirect contacts (i.e., 2-hop neighbors) to direct contacts (i.e., 1-hop neighbors).
For each network, we first compute the ratio of the number of indirect contacts transformed into direct contacts during the study period.
We call this ratio the transformation rate and denote it as $\TransRate$.
We then group respondents (i.e., ego nodes) with conditions of $\TransRate\,{=}\,0$ (i.e., no transformation) as \texttt{Class\,0}  and $\TransRate\,{\geq}\,0.2$ (i.e., high transformation rate) as \texttt{Class\,1}.
\texttt{Classes\,0} and \texttt{1} consist of 56 and 136 respondents, respectively.
We then classify these classes, resulting in accuracies of 1.0 and 0.82 for Steps 2 and 3.

\begin{figure}[tb]
    \centering
    \includegraphics[width=\linewidth]{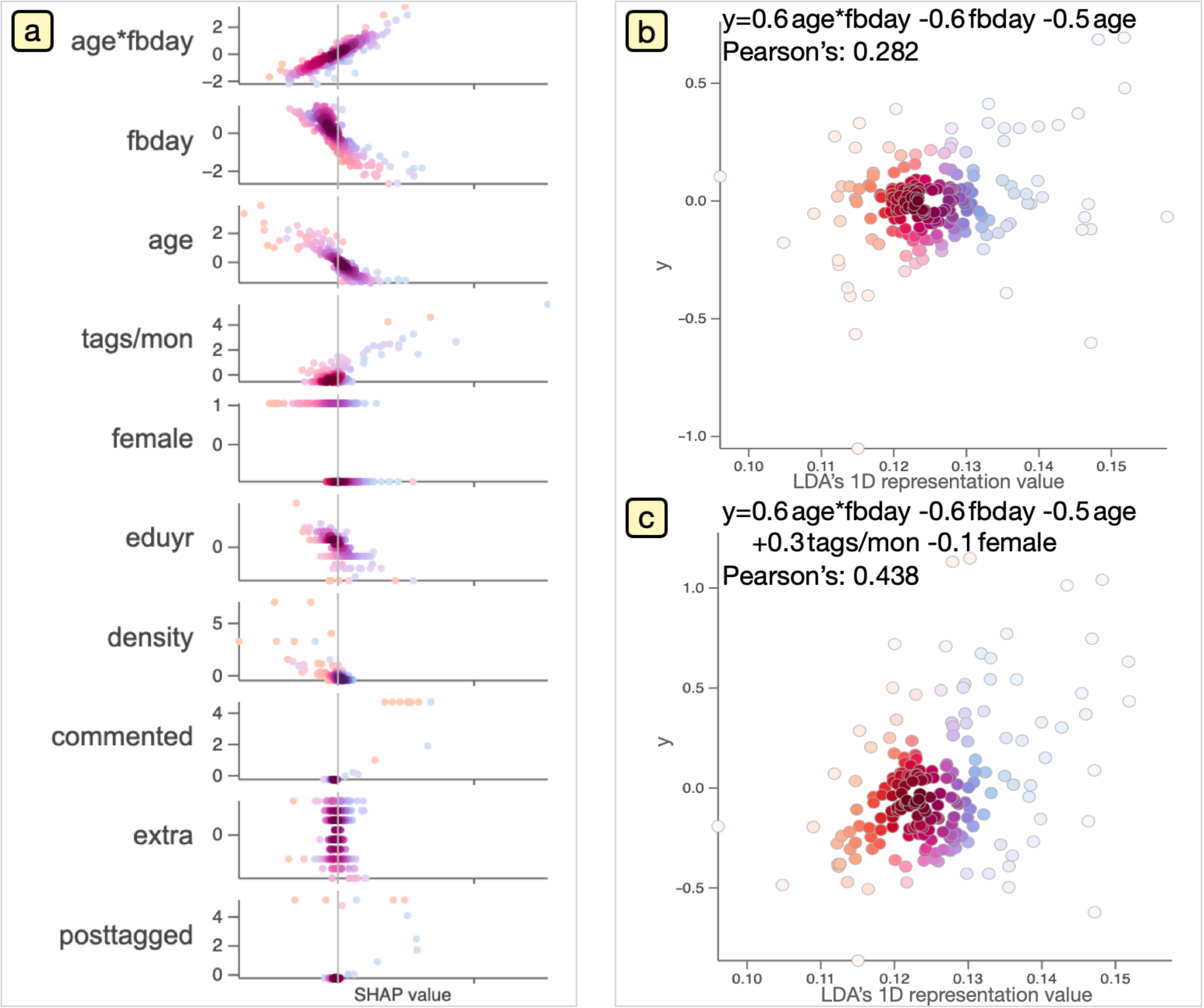}
    \caption{\CaptionCaseThree{}}
    \label{fig:case3}
\end{figure}

\autoref{fig:case3}-a shows the top 10 contributed attributes to the 1D representation. 
We see several clear trends between attributes' values and the SHAP values.
For example, \texttt{age*fbday} shows an increasing trend (i.e., larger \texttt{age*fbday} has a more positive impact on the transformation). 
\texttt{age*fbday} is the predefined composite variable by the existing study~\cite{lee2022indirect}.
This attribute is made by multiplying the ego's \texttt{age} and total days of Facebook use, \texttt{fbday}.
Due to this definition of \texttt{age*fbday}, all top 3 contributed attributes, \texttt{age*fbday}, \texttt{fbday}, \texttt{age}, are likely to interrelate with each other. 
Then, we construct a composite variable with these attributes.
As shown in \autoref{fig:case3}-b, the constructed composite variable, 0.6\,\texttt{age*fbday} -\,0.6\,\texttt{fbday} -\,0.5\,\texttt{age}, presents a weak correlation (Pearson's: 0.282). 
By investigating three attributes individually, we observe that they show a much weaker correlation to the 1D representation---\texttt{age*fbday}: 0.064, \texttt{fbday}: 0.052, \texttt{age}: 0.123. 
Thus, the composite variable seems to find more meaningful information by disentangling complex relationships among the three attributes.
We further construct a new composite variable by adding other highly ranked attributes, \texttt{tags/mon} (the average times of being tagged per month) and \texttt{female} (whether ego's gender is female or not).
As shown in \autoref{fig:case3}-c, the new composite variable has a significantly stronger correlation than the previous one (0.438~vs.~0.282).

For verification, we compare our findings with the existing study~\cite{lee2022indirect}.
The existing study employed a mixed-effect model to capture attributes' contributions to the transformation of indirect contacts. 
The top 5 contributed attributes suggested by their model are \texttt{female}, \texttt{extra}, \texttt{comments/mon} (the average number of comments made per month), \texttt{fbday}, and \texttt{age*fbday}. 
First of all, three attributes, \texttt{age*fbday}, \texttt{fbday}, and \texttt{female}, are seen in both their results and ours. 
Also, \texttt{tags/mon} and \texttt{comments/mon} should have semantic similarities with each other based on their meanings. 
In fact, by replacing \texttt{tags/mon} with \texttt{comments/mon}, we obtain the composite variable with 0.382 Pearson's correlation coefficient, which is considerably close to the original (Pearson's: 0.438).
These observations indicate that our NN-based model successfully captured similar information to the mixed-effect model.
The unique strength of our approach is in its ability to convey inter-relationships of attributes with their weights in the composite variable. 
While the existing work~\cite{lee2022indirect} crafted \texttt{age*fbday}, our study suggests (0.6\,\texttt{age*fbday} -\,0.6\,\texttt{fbday} -\,0.5\,\texttt{age}) or a more simplified version, (\texttt{age*fbday} -\,\texttt{fbday} -\,\texttt{age}), as a potential composite variable that better capture the influence on the transformation than \texttt{age*fbday}.

\section{Expert Feedback}

\noindent
To further validate our workflow’s usability, we conducted a focus group with five experts in social network studies.
The first expert (E1) is a distinguished researcher in an institute of sociology who collected Datasets I and II and also conducted research on these datasets with different focuses from ours. 
The second expert (E2) is an assistant professor in a department of sociology who formulated the guidelines for collecting the datasets.
The other three experts are researchers in institutes of sociology (E3) and statistical science (E4, E5) who also studied the same datasets.
The focus group was conducted through a video conference setup, where our workflow methods, visualization designs, and the three case studies were presented using an interactive demo combined with static screenshots.
Then, the five experts provided their comments as qualitative feedback on our workflow.

All the experts agreed that our workflow can support a wide range of their analysis targets as well as derive insights with more intuitive interpretations when compared to their current approach.
For example, E1 commented, ``The results [seen in the composite variables] are much easier to understand and intuitive than the outputs from the statistical models used in our previous research.''
All of them showed strong interest in insights that can be derived from the signed weights in composite variables.  
On the other hand, E1 noted a potential limitation of the datasets and analyses: ``The distribution of the respondents might affect the results related to the internet addiction because the students who granted the use of their data tended to be more addicted to the internet usage than others.''
E2 suggested better input attributes for the analyses, and we improved our case studies, accordingly.   

There were several discussions on our workflow usage and design.
E3 asked, ``What if there are attributes with a dominant influence on the result? What if the 1D representation does not have clear separation?'' 
For the former, the visualization shown in \autoref{fig:ui}-a can be used to identify dominant attributes and exclude them from inputs if such attributes exist.
Similarly, for the latter case, the visualization shown in \autoref{fig:ui}-b1 is useful to review the quality of the separation.
Then, based on the quality, analysts can take actions, such as adding more input attributes, reducing the ranges of the two ends of the output attribute, and improving the NN model.
E4 raised a question on our post-hoc simplification of network representations: ``Why not only use a neural network by designing the last layer with a linear activation function, instead of using LDA?''
Our answer is given in \autoref{sec:repr_simplification}. 
E5 asked, ``Why not use other centralities but these three?''
We selected degree, eigenvector, and betweenness centralities as a set of the most fundamental measures of structural characteristics.
Our workflow is flexible to employ any other centralities based on analysts' interests.

Finally, there were comments on our two-class density scatterplot design. 
E1 and E2 noted its analytical usability: ``The design of the two-class density scatterplot facilitates the identification of dense areas, providing better support for observing data distribution and correlations.''
E2 further added: ``We often wasted a considerable amount of time in validating hypotheses that are formed from misleading visual patterns of correlations and class separations in conventional scatterplots.''
E1 also commented: ``The design of the two-class density scatterplot can be particularly useful for large-scale data and has contributions to statistical science research.''
\section{Discussion}

\noindent
Through the case studies and the expert interview, we discussed the efficacy of our workflow and interactive visualizations. 
In Supplementary Materials~\cite{supp}, we further demonstrate the effectiveness of the constructed composite variables for conventional analyses, such as statistical hypothesis testing.
Besides these evaluations, to allow readers to test and evaluate our workflow and UI, we provide related source code as well as processed data~\cite{supp} from a publicly available dataset of faculty networks~\cite{wapman2022quantifying}.
Here, we provide additional discussions on our designs.

\textbf{Applicability to various data types.} 
We have designed our workflow for multivariate networks.
As multivariate networks have both high-dimensional and relational characteristics, by their nature, the workflow is even applicable to high-dimensional data and univariate network data.
For example, we can analyze high-dimensional data by skipping the extraction of network centrality-related attributes and the visualization of networks.
In addition, our workflow design can potentially adapt to advanced models of networks such as those containing meta-nodes and hyperedges. 
For example, as long as meta-nodes have the same set of node attributes as simple nodes in a network, we can still apply our workflow as is. 
Our workflow precomputes node- and link-related features before the training utilizing NNs; thus, we can deal with hyperedges during this preprocessing step (e.g., including hyperedges when applying relational functions in DeepGL).

\textbf{Other potential algorithm designs.}
While we employ the NNs, DR, and composite variable construction to support our analysis target, there are other potential designs.
One common way to understand the associations among a target and other attributes is applying a decision tree (DT)~\cite{brand2021uncovering,li2021visual}. 
When compared with a DT-based analysis, our design provides two main strengths in the interpretation step: simplicity and informativity.
A DT provides a set of attribute ranges that is useful to classify a target attribute. 
However, the number of ranges would be easily overwhelming when analyzing networks with many attributes. 
Also, from the DT results, it is difficult to numerically assess the influences from multiple attributes, such as those seen in the composite variables we have constructed for the case studies.

Another possible design is enabling the construction of more complicated composite variables, such as those with multiplications and logarithms of attributes.
Although this would be more effective to analyze complex relationships among attributes (e.g., \texttt{age*fbday} in \autoref{sec:cs3}), this construction requires much more complicated optimizations than ours.
One potential way to perform such advanced constructions while avoiding excessive computation is incorporating analysts' knowledge more actively (e.g., predetermining a part of a composite variable). 
We plan to investigate this direction in future research.

\textbf{Usability of two-class density scatterplots.}
We developed two-class density scatterplots to depict various patterns (e.g., distributions of class instances, trends, clusters, and outliers) in a single visualization.
The use of this scatterplot is not limited to the targeted analyses in this work. 
As binary classifications or group comparisons are frequently performed for ML and visual analytics, we believe two-class density scatterplots can contribute to these fields. 
As future research, we plan to evaluate our scatterplot design by conducting a comprehensive user study and designing new quality measures for multiclass scatterplots that convey both aggregated- and instance-level information (e.g., splatterplots~\cite{mayprga2013splatterplots} and winglets~\cite{lu2020winglets}).
This evaluation would also identify the two-class density scatterplot's shortcomings for further improvements.
There is one clear limitation: allowing visualization of only two classes. 
One potential way to support three or more classes is taking similar approaches to the multiclass geographical maps by Jo et al.~\cite{jo2018declarative}.
For example, we can first partition a 2D space based on the changes in the distribution of class instances and then display the distribution in each partitioned region with a bar-chart glyph.
We expect that this design can deal with several classes (e.g., five classes); however, this partition-based approach would not be suitable to reveal outliers and clusters.
Thus, we would like to investigate an extension to three or more classes in the future.

\section{Conclusion}

\noindent
We have introduced a new visual analytics workflow designed to help find associations from complex multivariate networks.
The workflow integrates neural-network-based representation learning,  composite variable construction, and interactive visualizations.
Benefiting from these components, the workflow generates expressive as well as interpretable analytical results. 
The design of workflow is also suited for analyzing other simpler types of data, such as a univariate network and high-dimensional data.
Thus, our work potentially contributes to a wide range of applications that involve analyses of large, complex data.

\section{Acknowledgments}
\noindent This work has been supported in part by the National Institute of Health through grants 1R01CA270454-01 and 1R01CA273058-01 and by the Knut and Alice Wallenberg Foundation through Grant KAW 2019.0024.

\bibliographystyle{IEEEtran}
\bibliography{00_ref}

\begin{IEEEbiography}[{\includegraphics[width=1in,height=1.25in,clip,keepaspectratio]{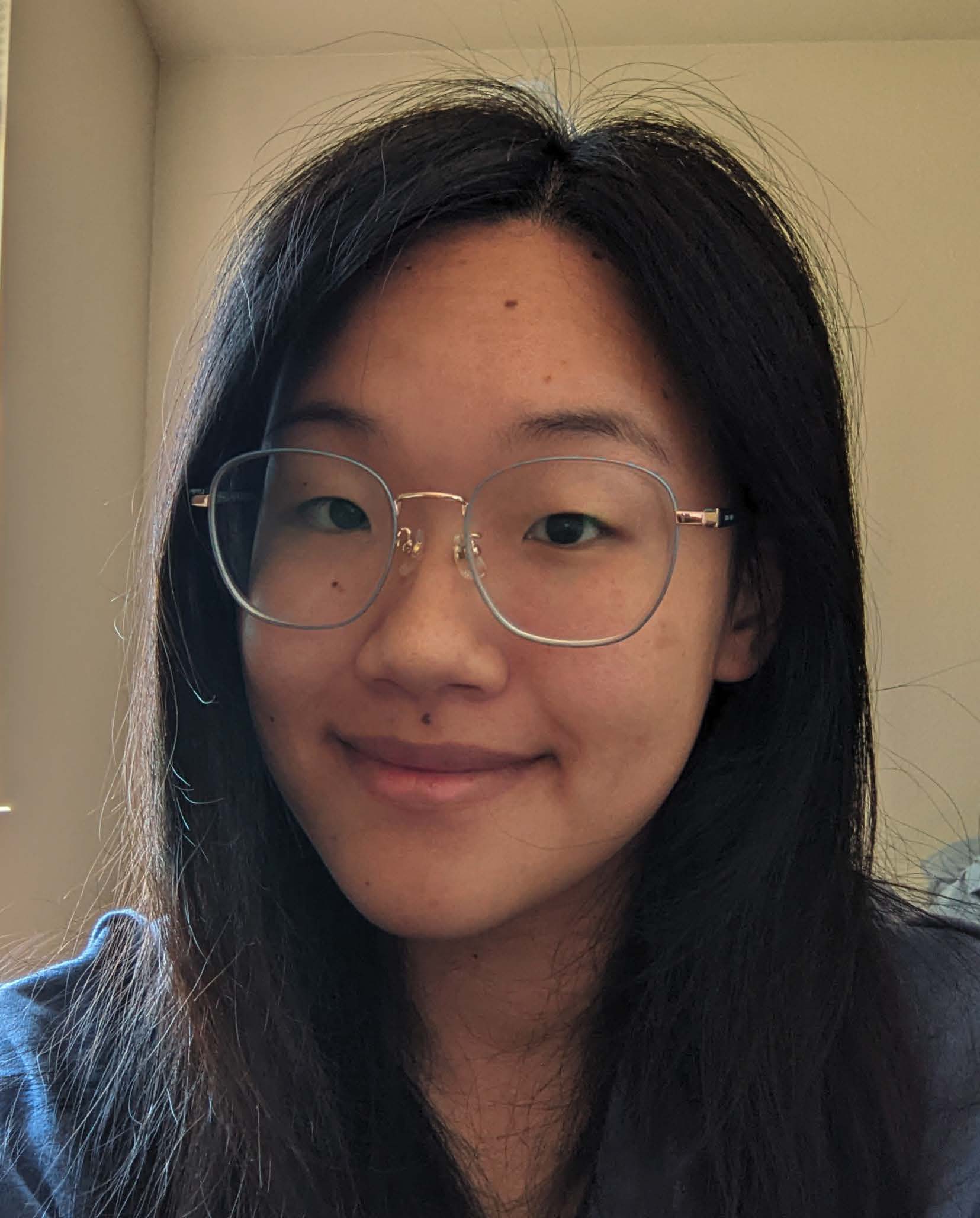}}]{Hsiao-Ying Lu} received the BS degree in Computer Science and Information Engineering from National Central University in Taiwan. Currently, she is pursuing a PhD degree at the University of California, Davis, focusing on Computer Science and Visualization under the guidance of Kwan-Liu Ma. Her research centers around integrating machine learning and visual analytics techniques to reason real-world complex networks.
\end{IEEEbiography}

\begin{IEEEbiography}[{\includegraphics[width=1in,height=1.25in,clip,keepaspectratio]{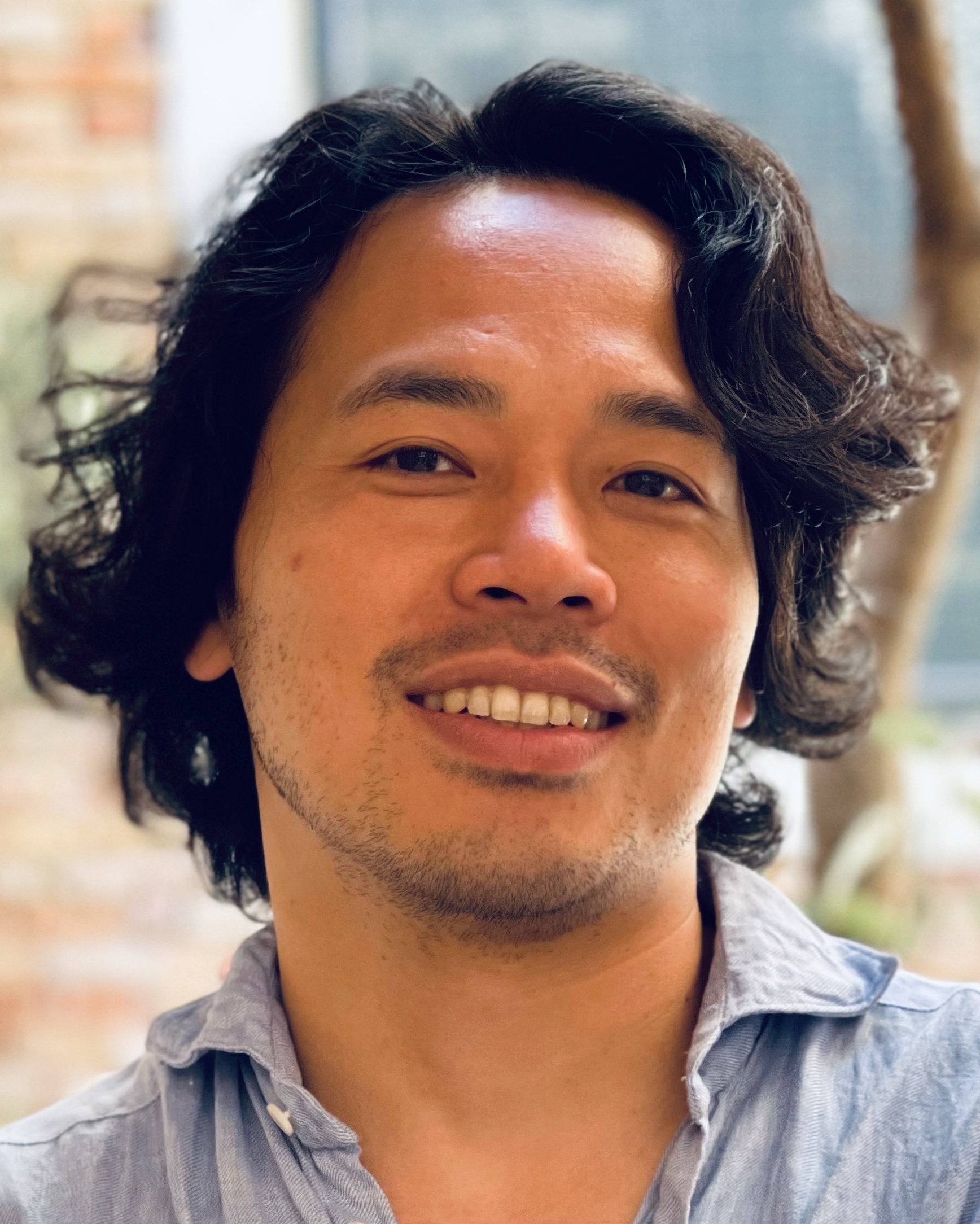}}]{Takanori Fujiwara} is a post-doctoral researcher at the Department of Science and Technology at Link\"oping University, Sweden. His expertise spans visual analytics, machine learning, and network science, and he specializes in developing interactive dimensionality reduction techniques. He publishes his research in top-tier visualization venues, such as the IEEE TVCG and the IEEE VIS conferences. His works have received Best Paper Honorable Mentions at the IEEE VIS (2019, 2023) and the IEEE PacificVis (2022) as well as the Best Graduate Researcher Award from the Department of Computer Science at UC Davis (2020). He received his Ph.D. degree in Computer Science from UC Davis and his Master's and B.E. from the University of Tokyo. Prior to his Ph.D., he worked for Kajima Corporation in Japan. 
\end{IEEEbiography}

\begin{IEEEbiography}[{\includegraphics[width=1in,height=1.25in,clip,keepaspectratio]{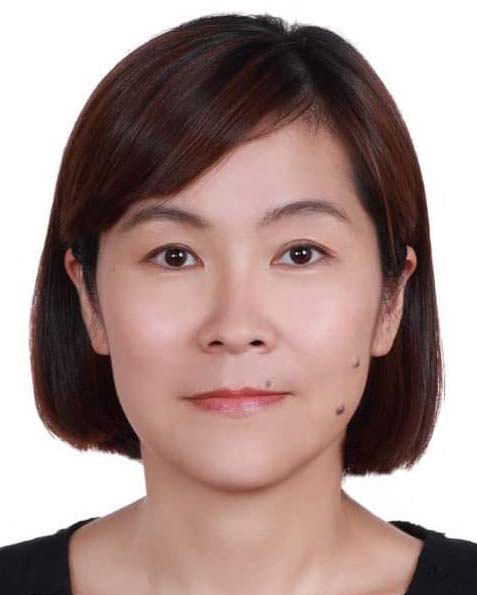}}]{Ming-Yi Chang} received the Ph.D. degree from the Department of Education, National Chengchi
University (NCCU), Taipei, Taiwan, in 2008. She is currently an Assistant Professor with the Department of Sociology, Fu Jen Catholic University (FJU), New Taipei City, Taiwan. Her research interests include social network analysis, sociology, and data mining.
\end{IEEEbiography}

\begin{IEEEbiography}[{\includegraphics[width=1in,height=1.25in,clip,keepaspectratio]{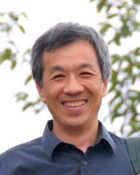}}]{Yang-chih Fu} is a Distinguished Research Fellow at the Institute of Sociology, Academia Sinica, Taiwan. He received the Ph.D. in Sociology from the University of Chicago. His recent research, which involves collecting contact records on social media and self-reported contact diaries from diverse participant groups, examines how social outcomes vary at the network, individual, tie, and contact levels. The findings and interpretations from these contact-based personal network studies have primarily been published in the journal Social Networks. Prof. Fu was awarded the Karl Polanyi Prize for his collaborative research comparing contact networks between Taiwan and Hungary.
\end{IEEEbiography}

\begin{IEEEbiography}[{\includegraphics[width=1in,height=1.25in,clip,keepaspectratio]{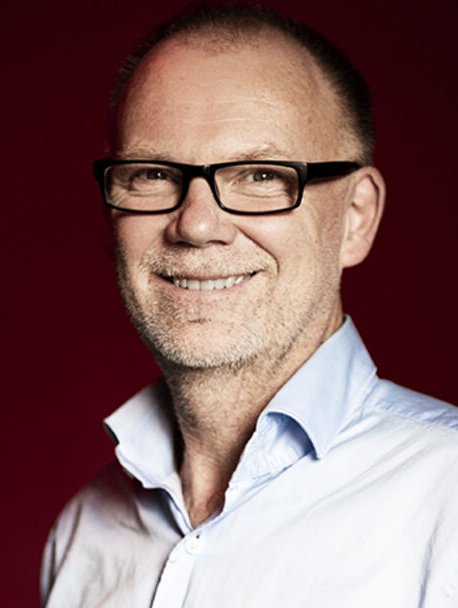}}]{Anders Ynnerman} received a Ph.D. 1992 in physics from Gothenburg University. During the early 90’s he was doing research at Oxford University, UK, and Vanderbilt University, USA. In 1996 he started the Swedish National Graduate School in Scientific Computing, which he directed until 1999. From 1997 to 2002 he directed the Swedish National Supercomputer Centre and from 2002 to 2006 he directed the Swedish National Infrastructure for Computing (SNIC). Since 1999 he is holding a chair in scientific visualization at Linkoping University and he is the director of the Norrkoping Visualization Center--C, which currently constitutes one of the main focal points for research and education in computer graphics and visualization in the Nordic region. The center also hosts a public arena with large scale visualization facilities. He is also one of the co-founders of the Center for Medical Image Science and Visualization (CMIV).
\end{IEEEbiography}

\begin{IEEEbiography}[{\includegraphics[width=1in,height=1.25in,clip,keepaspectratio]{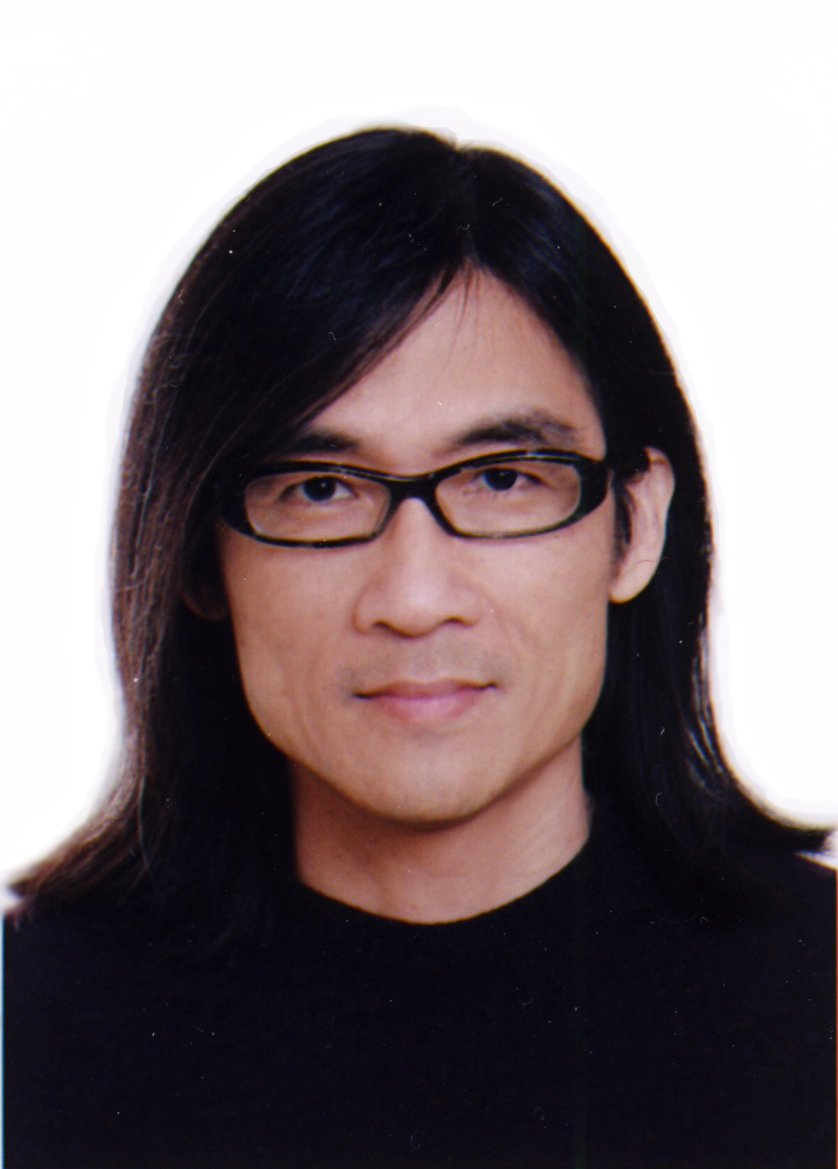}}]{Kwan-Liu Ma} is a distinguished professor of computer science at the University of California, Davis, where he leads VIDI Research Group. Professor Ma received his PhD degree in computer science from the University of Utah in 1993. His research interests include visualization, computer graphics, human computer interaction, and machine learning. For his significant research accomplishments, Professor Ma has received many recognitions, including the NSF PECASE award in 2000, IEEE Fellow 2012, 
the IEEE VGTC Visualization Technical Achievement Award in 2013, the IEEE Visualization Academy in 2019, 
and ACM Fellow in 2023. He has served as papers co-chair for SciVis, InfoVis, EuroVis, PacificVis, Graph Drawing, and on the editorial board of IEEE TVCG (2007-2011) and IEEE CG\&A (2007-2019), ACM TiiS (2021-present). Professor Ma presently serves on both the IEEE Visualization Steering Committee and IEEE PacificVis Steering Committee. 
\end{IEEEbiography}

\vfill

\end{document}